\documentclass[a4paper,twocolumn,english,showpacs,secnumarabic,aps,pre]{revtex4}

\usepackage{textcomp}

\usepackage{amssymb}
\usepackage{amsmath}
\usepackage{amsfonts}

\usepackage{float}

\usepackage{graphicx}
\usepackage{graphics}

\usepackage[colorlinks=true,urlcolor=blue,citecolor=blue]{hyperref}

\usepackage{babel}

\makeatletter

\newcommand{\noun}[1]{\textsc{#1}}
\providecommand{\tabularnewline}{\\}

\makeatother

\begin{document}

\title{Master crossover behavior of parachor correlations for one-component
fluids.}

\author{Yves Garrabos, Fabien Palencia, Carole Lecoutre }

\affiliation{Equipe du Supercritique pour l'Environnement, les Matériaux et l'Espace,
ICMCB-CNRS UPR 9048, Université Bordeaux I, 87 avenue du Dr A. Schweitzer,
F 33608 PESSAC Cedex France.}

\author{Daniel Broseta}

\affiliation{Laboratoire des Fluides Complexes, UMR 5150 - Université de Pau et
des Pays de l'Adour, B.P. 1155, 64013 Pau Cedex, France.}

\author{Bernard Le Neindre}

\affiliation{Laboratoire des Interactions Moléculaires et des Hautes Pressions,
UPR 1311 - Centre National de la Recherche Scientifique - Université
Paris XIII , avenue Jean Baptiste Clément, F 93430 Villetaneuse, France.}

\date{09 July 2007}

\pacs{64.60.Ak.;05.10.Cc.;05.70.Jk;65.20.+w}

\begin{abstract}
The master asymptotic behavior of the usual parachor correlations,
expressing surface tension $\sigma$ as a power law of the density
difference $\rho_{L}-\rho_{V}$ between coexisting liquid and vapor,
is analyzed for a series of pure compounds close to their liquid-vapor
critical point, using only four critical parameters $\left(\beta_{c}\right)^{-1}$,
$\alpha_{c}$, $Z_{c}$ and $Y_{c}$, for each fluid. This is accomplished
by the scale dilatation method of the fluid variables where, in addition
to the energy unit $\left(\beta_{c}\right)^{-1}$ and the length unit
$\alpha_{c}$, the dimensionless numbers $Z_{c}$ and $Y_{c}$ are
the characteristic scale factors of the ordering field along the critical
isotherm, and the temperature field along the critical isochor, respectively.
The scale dilatation method is then formally analogous to the basic
system-dependent formulation of the renormalization theory. Accounting
for the hyperscaling law $\frac{\delta-1}{\delta+1}=\frac{\eta-2}{2d}$,
we show that the Ising-like asymptotic value $\pi_{a}$ of the parachor
exponent is unequivocally linked to the critical exponents $\eta$,
or $\delta$, by $\frac{\pi_{a}}{d-1}=\frac{2}{d-\left(2-\eta\right)}=\frac{\delta+1}{d}$
(here $d=3$ is the space dimension). Such \emph{mixed} hyperscaling
laws combine either the exponent $\eta$, or the exponent $\delta$
which characterizes \emph{bulk} critical properties of $d$ dimension
along the critical isotherm or exactly at the critical point, with
the parachor exponent $\pi_{a}$ which characterizes interfacial properties
of $d-1$ dimension in the non-homogeneous domain. Then we show that
the asymptotic (symmetric) power law $\left(\alpha_{c}\right)^{d-1}\beta_{c}\sigma=D_{\rho}^{\sigma}\left(\frac{\rho_{L}-\rho_{V}}{2\rho_{c}}\right)^{\pi_{a}}$
is the two-dimensional critical equation of state of the liquid-gas
interface between the two-phase system at constant total (critical)
density $\rho_{c}$. This power law complements the asymptotic (antisymmetric)
form $\left(\mu_{\rho}-\mu_{\rho,c}\right)\frac{\rho_{c}}{p_{c}}=\pm D_{\rho}^{c}\left|\frac{\rho-\rho_{c}}{\rho_{c}}\right|^{\delta}$
of the three-dimensional critical equation of state for a fluid of
density $\rho\neq\rho_{c}$ and pressure $p\neq p_{c}$, maintained
at constant (critical) temperature $T=T_{c}$ {[}$\mu_{\rho}$ ($\mu_{\rho,c}$)
is the specific (critical) chemical potential; $p_{c}$ is the critical
pressure; $T_{c}$ is the critical temperature]. We demonstrate the
existence of the related universal amplitude combination $D_{\rho}^{c}\left(D_{\rho}^{\sigma}\right)^{\frac{d}{1-d}}=R_{D\sigma}=\text{universal}\,\mbox{constant}$,
constructed with the amplitudes $D_{\rho}^{\sigma}$ and $D_{\rho}^{c}$,
separating then the respective contributions of each scale factor
$Y_{c}$ and $Z_{c}$, characteristic of each thermodynamic path,
i.e., the critical isochore and the critical isotherm (or the critical
point), respectively. The main consequences of these theoretical estimations
are discussed in the light of engineering applications and process
simulations where parachor correlations constitute one of the most
practical method for estimating surface tension from density and capillary
rise measurements.
\end{abstract}
\maketitle

\section{Introduction}

Most of the phenomenological approaches for modelling the fluid properties
in engineering applications are commonly based on the extended corresponding-states
principle \citet{Poling2001}. In this scheme, the estimation of thermodynamic
properties can be made using multiparameter equations of state, that
account for increasing molecular complexity by increasing the number
of adjustable parameters. Such \emph{engineering} equations of state
(whose mathematical forms must be compatible for practical use in
fluid mixture cases), are then generally convenient tools to estimate
a single phase property with sufficient accuracy \citet{Poling2001}.
However the knowledge of properties in the nonhomogeneous domain \citet{Rowlinson 1984},
such as the surface tension $\sigma$, the capillary length $\ell_{Ca}$,
the density difference $\Delta\rho_{LV}=\rho_{L}-\rho_{V}$ between
the coexisting liquid and vapor phases of respective density $\rho_{L}$
and $\rho_{V}$, are also of prime importance to gain confidence in
fluid modelling and process simulations (geological fluid flows, assisted
recovery of oil, storage of green house gases, pool boiling phenomena,
microfluidic devices based on wetting phenomena, etc). Therefore,
a large number of related phenomenological laws, referred to as \emph{ancillary}
equations, have been proposed in the literature \citet{Poling2001,Rowlinson 1984,Xiang2005}
to calculate such properties in the nonhomogeneous domain. This complementary
approach generally leads to unsolvable mathematical differences with
values calculated from the equations of state, and increases in a
substantial manner the number of adjustable parameters to account
for complex molecular fluids.

Before focussing on the specific form of ancillary equations between
$\sigma$ and $\Delta\rho_{LV}$ \citet{Macleod1923}, the so-called
parachor correlations \citet{Poling2001,Broseta2005}, it is interesting
to recall the two well-known practical interests of the fluid modelling
based on the extended corresponding-states principle: i) the thermodynamics
properties of a selected pure fluid are fully specified from a few
fluid-dependent parameters such as for example its critical coordinates
$T_{c}$ (critical temperature), $p_{c}$ (critical pressure) and
$\bar{v}_{c}$ (critical molar volume) in the original and simplest
form of the corresponding-states principle \citet{Guggenheim1945};
ii) the most convenient tools to estimate the fluid phase surface,
including the two-phase equilibrium lines, are provided by the cubic
and generalized van der Waals equations \citet{Poling2001}. Our main
objective in this introductive discussion is then only to recall the
number and the nature of the most usefull macroscopic parameters used
in such engeneering equations of state (for a review see Ref. \citet{Anderko2000}).
For more detailed presentations of the basic understanding from a
rigorous microscopic approach of the molecular interaction and the
theoretical background for developping better functional forms of
the pressure-volume relationship see for example the Refs. \citet{Hirschfelder1964,Rowlinson1971,Hansen 1986,Poling2001}
and the review of Ref. \citet{Ely2000}.

It was well-known \citet{Guggenheim1945} that only the inert gases
(Ar, Kr, Xe) can obey the two-parameter corresponding-states principle
 (i.e., an energy unit and a length unit mandatorily needed to compare
dimensionless thermodynamic states for same values of the dimensionless
independent variables, admitting that the molar mass of each one-component
fluid is known). This restrictive conclusion was founded on results
obtained in building unique functions of the reduced thermodynamic
variables, examinating many thermodynamics properties, such as the
density difference between coexisting liquid and vapour phases, the
saturated vapour pressure curve, the second virial coefficient, etc.

Considering the modelling based on statistical mechanics \citet{Hirschfelder1964,Rowlinson1971,Hansen 1986},
this two-parameter corresponding-states description can be validated
from the restrictive compounds made of spherical atoms with centro-symmetrical
forces (such as precisely the inert gases mentioned above). The short-ranged
space ($r$) dependence of intermolecular pair potentials $u\left(r\right)$\citet{Maitland1981}
can be written for example in the form $u\left(r\right)=\epsilon_{\text{LJ}}F^{\text{LJ}}\left(\frac{r}{\sigma_{\text{LJ}}}\right)$
where $F^{\text{LJ}}$ is the Lennard Jones (12-6) universal function
\citet{Hirschfelder1964}. The two quantities $\epsilon_{\text{LJ}}$
and $\sigma_{\text{LJ}}$ are scaling (energy and length) parameters
which characterize a particular substance. Compounds which obey this
kind of universal potential function with two microscopic scaling
parameters are said to be conformal \citet{Rowlinson1971,Hansen 1986,Poling2001}.

On the other hand, some intermolecular potential models with attractive
interaction forces of infinite range, have given physical reality
to the famous form (cubic with respect to volume) of the van der Waals
(vdW) equation of state \citet{vanderWaals1973}, separating then
the repulsive and attractive contribution to the pressure-volume relationship
estimated from the generalized van der Waals theory \citet{Anderko2000}.
Although the two pressure terms of the original van der Waals equation
do not quantitatively represent the true repulsive and attractive
forces, the introduction of the two characteristic constants for each
fluid - its actual covolume $b$, not available to molecular motion
due to a finite diameter of each repulsive molecule, and the amplitude
$a$ of the pressure decrease due to the intermolecular attraction
-, has proven to be extremely valuable for the representation of its
properties. Thus, after expressing the values of the van der Waals
parameters $a$ and $b$ at the critical point, the unique function
$\frac{p}{p_{c}}=f^{\text{vdW}}\left(\frac{T}{T_{c}},\frac{\bar{v}}{\bar{v}_{c}}\right)$
of the original van der Waals equation conforms to the two-parameter
corresponding-states principle since $\bar{v}_{c}$ depends unequivocally
on $T_{c}$ and $p_{c}$, through the unique value of the critical
compression factor $Z_{0c}^{\text{vdW}}=\frac{p_{c}\bar{v}_{c}}{RT_{c}}=\frac{3}{8}$. 

So that, at the macroscopic level, practical formulations of the two-parameter
corresponding-states principle employ as scaling parameters the critical
temperature $T_{c}$ (providing energy unit by introducing the Boltzman
factor $k_{B}$), and the critical pressure $p_{c}$ (providing a
length unit through the quantity $\left(\frac{k_{B}T_{c}}{p_{c}}\right)^{\frac{1}{d}}$
expressed for space dimension $d=3$), and seek to represent thermodynamic
properties, thermodynamic potentials and related equations of state
as universal (i.e., unique) dimensionless functions of the new reduced
variables $\frac{T}{T_{c}}$ and $\frac{p}{p_{c}}$ (or $\frac{\bar{v}}{\bar{v}_{c}}$).
However, although this principle only applies to conformal fluids,
it is easy to show that it allways generates unreductible difficulties
to obtain satisfactorily agreement between theoretical modelling and
experimental results, especially for the two-phase surface approaching
the liquid-gas critical point. For example, the potential parameters
$\epsilon_{\text{LJ}}$ and $\sigma_{\text{LJ}}$ of a Lennard-Jones
(12-6) fluid evaluated from different thermodynamic and transport
properties of the same real fluid tend to be significantly different
than the ones directly obtained from their relations to the critical
point coordinates (although, according to the molecular theory, the
calculated critical compression factor $Z_{0c}^{\text{LJ}}\simeq0.290$
remains the same for all these conformal fluids). Moreover, real atoms
like Ar, Kr, and Xe are definitively not conformal ($Z_{c}$, $\alpha_{c,R}$,
and then $Y_{c}$, $\frac{p_{TP}}{p_{c}}$, etc., are not strictly
constant numbers \citet{Rowlinson1971}). Similarly, the well-known
breacking up of van der Waals equation of state occurs immediately,
noting that the value $Z_{0c}^{\text{vdW}}=0.375$ significantly differs
from the $Z_{c}$ values of real fluids {[}ranging for example from
$Z_{c}\left(\text{H}_{2}\text{O}\right)\simeq0.22$ to $Z_{c}\left(^{4}\text{He}\right)\simeq0.30$],
especially in the inert gase case {[}for example $Z_{c}\left(\text{Xe}\right)=0.286$]. 

As stated above in the developpement of its simplest form from fluid
state theories, a two-parameter description does not hold for real
atoms and \emph{a fortiori} for molecules with more complex shapes
and interactions. Indeed, for compounds with non associating and non
(or weakly) polar interactions of nonspherical molecules, also referred
to as normal compounds, the deviations from the two-parameter corresponding-states
modelling were most often described by one additional parameter, the
so-called acentric factor, $\omega=-1-\log\left[\frac{p_{\text{sat}}\left(T=0.7\, T_{c}\right)}{p_{c}}\right]$,
proposed by Pitzer \citet{Pitzer1955}. The acentric factor was defined
from the reduced value $p^{*}=\frac{p_{\text{sat}}}{p_{c}}$ of the
saturated vapor pressure $p_{\text{sat}}\left(T\right)$ at the reduced
value $T^{*}=\frac{T}{T_{c}}=0.7$ of the vapor saturation temperature,
such that it is essentially $\omega\approxeq0$ for inert gases Ar,
Kr, and Xe. An inert vapor condensates at one tenth of the critical
pressure at $T^{*}=0.7$, while a vapour of more complex molecules
condensates at lower relative pressure, leading to $\omega$ positive,
so that, the larger and more elongated the molecule, the larger $\omega$,
due to an increasing contribution of the attractive molecular interaction.
Thermodynamics properties of the normal compounds can then be described
by unique functions of the three parameters $T_{c}$, $p_{c}$, $\omega$.

It is also well-established \citet{Anderko2000,Poling2001} that this
three-parameter coresponding state modelling can be accounted for
by using a three-parameter equation of state, for example the frequently
referred Soave-Redlich-Kwong \citet{Redlich1949,Soave1972}, Peng-Robinson
\citet{Peng1976}, or Patel-Teja \citet{Patel1982,Patel1996} cubic
equations of state. When a third parameter is introduced into a cubic
equation of state, the critical compression factor $Z_{0c}$ becomes
fluid-dependent, as stated for real fluids. Unfortunately, although
a three-parameter equation can be forced to the correct $Z_{c}$,
only better overall improvement of the accuracy in estimations of
the phase surface is obtained when its calculated value is greater
than the real one (for example $Z_{0c}^{\text{RK}}=0.333$ and $Z_{0c}^{\text{PR}}=0.307$
for Redlich-Kwong (RK) and Peng-Robinson (PR) equations of state,
respectively, while Patel-Teja (PT) equation of state treats this
calculated compression factor $Z_{0c}^{\text{PT}}$ as an adjustable
parameter). More generally, such quantitatively inaccurate calculations
are due to the relative rigidity of the cubic form (which limits the
quality of the representation of derivative properties), added to
fundamental limitations of analytic equations close to the critical
point (which generate mean field behaviors of fluid properties).

Moreover, useful precise measurement of the saturation pressure curve
$p_{\text{sat}}\left(T\right)$, introduces a critical limiting (dimensionless)
slope $\alpha_{c,R}=\left[\frac{\partial\log\left(p_{\text{sat}}\right)}{\partial\log\left(T\right)}\right]_{T=T_{c}}=\frac{T_{c}}{p_{c}}\left(\frac{\partial\text{p}_{sat}}{\partial T}\right)_{T=T_{c}}$
at $T_{c}$, as another fluid characteristic parameter, also known
as the Riedel factor \citet{Riedel1954}. Anticipating then result
of the next Section which introduces the critical number $Y_{c}=\frac{T_{c}}{p_{c}}\left(\frac{\partial p_{\text{sat}}}{\partial T}\right)_{T=T_{c}}-1$
{[}see Eq. (\ref{isochoric factor (19)})], we note the relation $\alpha_{c,R}=Y_{c}+1$
between the Riedel factor and the dimensionless number $Y_{c}$. Therefore,
as an immediate consequence of the real location of the liquid-gas
critical point in the experimental $p,\bar{v},T$ phase surface, the
addition of $Z_{c}$ and $\alpha_{c,R}$ (or $Y_{c}$) to $\omega$
appears as a useful parameter set increment, able to describe deviations
from the two-parameter corresponding-states principle based on $T_{c}$,
$p_{c}$ knowledge. Obviously, any three-parameter corresponding-states
modelling needs implicit dependence between $\omega$, $Z_{c}$, and
$\alpha_{c,R}$ (or $Y_{c}$), which provides bases for a large number
of three-parameter corresponding-states models by developping empirical
combinations between $\omega$, $Z_{c}$, and $\alpha_{c,R}$ (or
$Y_{c}$), such as ones where $Z_{c}$ is linearly related to $\omega$
\citet{Schreiber1989} for normal compounds. This constrained situation
to reproduce the critical point location in the $p,T$ diagram, is
certainly the most important practical reason why the three-parameter
cubic equations of state (which allow a fair thermodynamic description
of the normal compounds \citet{Poling2001}, including their interfacial
properties \citet{Miqueu2000}), are the most popular equations of
state developped again today for industrial process design.

However, the three-parameter corresponding-states modelling still
remains not appropriate for describing highly polar and {}``associating''
fluids (such as water or alcohols for example). At least an additional
fourth parameter is needed to extend the corresponding-states approaches,
which leads to multiple routes to account for this increasing complexity
of the microscopic molecular interaction. Several empirical expressions
have been proposed for this fourth parameter increment, such as the
one introducing the Stiel polar factor \citet{Halm1967} for example.
Again a myriad of four-parameter corresponding-states models can then
be defined using $T_{c}$, $p_{c}$, and practical combinations which
provide only two independent dimensionless numbers chosen among the
critical compression factor, the Pitzer acentric factor, the Riedel
factor, the Stiel polar factor, etc. For example, Xiang \citet{Xiang2005},
noticing that polar and nonpolar coumpounds may have similar $\omega$,
but different $Z_{c}$, or, in other words, that the relation between
$Z_{c}$ and $\omega$, that holds for normal fluids, does not hold
for polar and associated fluids, has proposed recently to use the
four parameters $T_{c}$, $p_{c}$, $Z_{c}$, and $\omega$. This
latter description can be then accounted for by using a four-parameter
equation of state, but noting that the results obtained from the four-parameter
equations which are constrained to reproduce the critical point, are
only slightly better than those obtained from the three-parameter
equations of state.

A notable exception, recently proposed by Kiselev and Ely \citet{Kiselev2003},
is the empirical implementation of the one-parameter (represented
by the Ginzburg number $Gi$ \citet{Ginzburg1960}) crossover description
in a generalized corresponding-states model which use the (four-parameter)
Patel-Teja equation of state $\frac{p}{p_{0c}^{\text{PT}}}=f^{\text{PT}}\left(\frac{T}{T_{c}},\frac{\bar{v}}{\bar{v}_{c}};\omega,Z_{0c}^{\text{PT}}\right)$
to calculate the classical behavior of the Helmholtz free energy far
away from the critical point (with the condition $\frac{p_{0c}^{\text{PT}}\bar{v}_{c}}{RT_{c}}=Z_{0c}^{\text{PT}}\leq\frac{1}{3}$).
In that approach, the experimental value of the critical molar volume
$\bar{v}_{c}$ replaces $\frac{k_{B}T_{c}}{p_{c}}$ as volume unit,
leading to a calculated critical pressure $p_{0c}^{\text{PT}}=Z_{0c}^{\text{PT}}\frac{RT_{c}}{\bar{v}_{c}}$
from an empirical correlation expressing the calculated critical compression
factor $Z_{0c}^{\text{PT}}\left(\omega,Z_{c}\right)$ as a function
of real acentric factor $\omega$ and real critical compression factor
$Z_{c}$. A redefinition of the generalized corresponding-states principle
in the form $\frac{p}{p_{0c}^{\text{PT}}}=f^{\text{CR}}\left(\frac{T}{T_{c}},\frac{\bar{v}}{\bar{v}_{c}};\omega,Z_{c},Gi\right)$
only introduces $Gi$ as an additional corresponding-states parameter.
Hereabove $f^{\text{CR}}$ is a unique function, which accounts for
a phenomenological crossover model that incorporates singular behavior
in the critical region, and transforms into an analytical equation
of state far away from the critical point. Thus, it was assumed that
the Ginzburg number $Gi\left(\omega,Z_{c},M_{\text{mol}}\right)$
can also be expressed as a function of real acentric factor $\omega$,
critical compression factor $Z_{c}$, and molar mass $M_{\text{mol}}$
of the fluid. As a final result, the unique four-parameter crosssover
equation $\frac{p}{p_{0c}^{\text{PT}}}=f^{\text{CR}}\left(\frac{T}{T_{c}},\frac{\bar{v}}{\bar{v}_{c}};\omega,Z_{c}\right)$,
similar to the classical four-parameter equation of state, is able
to predict with acceptable accuracy the phase surface near to and
far from the critical point, only using the real critical parameters
$T_{c}$, $\bar{v}_{c}$, $Z_{c}$, and the acentric factor $\omega$.
However, in such an empirical modelling, the real $p_{c}$ and $\left(\frac{\partial p_{\text{sat}}}{\partial T}\right)_{T=T_{c}}$can
never be accounted for as entry parameters which characterize the
one-component fluid. 

>From this brief status of the extended corresponding-states principle,
it seems undeniable that a minimum set made of four parameters is
necessary to characterize each one-component fluid, thus identifying
among a large set of usefull fluid-dependent parameters the four critical
parameters $T_{c}$, $p_{c}$, $Z_{c}$, and $\alpha_{c,R}$ (or $Y_{c}$).
If the introduction of $T_{c}$, $p_{c}$, $\bar{v}_{c}$, seems the
natural way to define the energy and length units and a related characteristic
critical compression factor, the arbitrary choice of the fourth parameter
confers empirical nature to any extended corresponding-states approach
(and to any functional form of the equation of state based on, especially
when the calculated compression factor differs from the real one).

It is then remarkable that the minimal set made of the four critical
parameters $T_{c}$, $p_{c}$, $Z_{c}$, and $Y_{c}$ could be alternatively
identified from a phenomenological analysis \citet{Garrabos1982,Garrabos1985,Garrabos1986,Garrabos2002,Garrabos2006qe}
of the singular behaviour of these fluids approaching their liquid-gas
critical point. Moreoever, a fundamental distinction occurs in this
critical phenomena scheme since the introduction of the two dimensionless
critical numbers $Z_{c}$ and $Y_{c}$ underlines their respective
asymptotic scale-factor nature (see below and Refs. \citet{Garrabos1982,Garrabos1985}).
As an essential consequence, the two-scale factor universality \citet{Privman1991}
estimated by the renormalization group method of field theory \citet{ZinnJustin2002}
is thus accounted for in building the \emph{master} (i.e., unique)
dimensionless crossover functions \citet{Bagnuls2002,Garrabos2006gb,Garrabos2006mcf}
of the \emph{renormalized} (i.e., rescaled) dimensionless field variables
\citet{Bagnuls1984a,Bagnuls1984b}. 

How to account for all these (practical and theoretical) results to
develop the appropriate forms of the ancillary equations remains a
difficult task, which was the object of few basic studies in regards
the practical importance of the two-phase fluid properties. For example,
the use of parachor correlations between $\sigma$ and $\Delta\rho_{LV}$
are a convenient way of estimating surface tension from density measurements.
Moreover, recent model calculations \citet{Giessen1999} and phenomenological
estimations \citet{Miqueu2001} have shown the strength of the parachor
correlations in the case of fluid mixtures.

For pure fluids, parachor correlations have the following form \begin{equation}
\sigma=\left(\frac{P_{e}}{M_{\text{mol}}}\Delta\rho_{LV}\right)^{\pi_{a,e}},\label{parachor correlation (1)}\end{equation}
where $M_{\text{mol}}=N_{A}m_{\bar{p}}$. $N_{A}$ is Avogadro number
and $m_{\bar{p}}$ is the molecular mass. The subscript $\bar{p}$
refers to a molecular property, i.e. a property of the constitutive
particle (atoms or molecules). The amplitude $P_{e}$, called \emph{parachor},
is a fluid-dependent property, while the parachor exponent $\pi_{a,e}$
is expected to have a unique numerical value for all the fluids. For
a review of the $\pi_{a,e}$ and $P_{e}$ values, see for example
Ref. \citet{Broseta2005}. A subscript $e$ indicates an \emph{effective}
value which corresponds to a given \emph{finite} experimental range
along the vapor-liquid equilibrium (VLE) line. Indeed, the introduction
in the early 1920's of Eq. (\ref{parachor correlation (1)}) was based
on experimental observations close to the triple point where $\rho_{V}\approx0\ll\rho_{L}$
(and then $\Delta\rho_{LV}\approx\rho_{L}$), leading to the first
proposed value of $\pi_{a,e}=4$ for the effective parachor exponent
\citet{Macleod1923}. It was then noted that the fluid parachor $P_{e}$
is approximately a constant value in a large (in absolute scale) temperature
range, leading to various attempts for its estimation by methods issued
from group contribution methods, or extended corresponding-states
principle. Accordingly, the parachor must be (at least) related to
the four parameters involved in engineering equations of state. 

However, it is now well-established \citet{Anisimov2000} that the
validity range of such \emph{scaling form} (\ref{parachor correlation (1)})
is \emph{strictly} restricted to the asymptotic approach of the liquid-gas
critical point, where $\sigma$ and $\Delta\rho_{LV}$ simultaneously
go to zero \citet{Widom1965,Fisk1969} with the universal features
of the uniaxial 3D Ising-like systems \citet{Privman1991,ZinnJustin2002}.
Indeed, in that liquid-gas critical domain, it is expected that $\Delta\rho_{LV}$
and $\sigma$ behave as $\Delta\rho_{LV}\sim\left(T_{c}-T\right)^{\beta}$
and $\sigma\sim\left(T_{c}-T\right)^{\phi}$, respectively, where
the critical exponents $\beta$ and $\phi$ take the following universal
values at $d=3$: $\beta\approxeq0.326$ \citet{Guida1998} and $\phi=\left(d-1\right)\nu\approxeq1.26$
\citet{Widom1965,Guida1998}. To obtain the latter scaling law, use
has been made of $\sigma\sim\xi^{d-1}$ \citet{Widom1965,Fisk1969},
where the power law $\xi\sim\left(T_{c}-T\right)^{-\nu}$ (with $\nu\approxeq0.630$
\citet{Guida1998}) accounts for the asymptotic singular behavior
of the correlation length $\xi$. Obviously, the asymptotic Ising-like
form of Eq. (\ref{parachor correlation (1)}) reads $\sigma\propto\left(\Delta\rho_{LV}\right)^{\frac{\phi}{\beta}}$,
with $\pi_{a,e}\rightarrow\pi_{a}=\frac{\phi}{\beta}\approxeq3.87$
when $\sigma\rightarrow0$ and $\Delta\rho_{LV}\rightarrow0$, i.e.,
a universal power law behavior with universal critical exponent $\pi_{a}$
of value which differs significantly from $\pi_{a,e}=4$ for example.

Therefore, despite the fact that the parachor correlations were initially
developed to correlate the surface tension of liquids close to their
triple point, their theoretical justification is only well-understood
close to the critical point. That infers a paradoxical situation when
the main objective is to estimate a single value of $P_{e}$ in the
\emph{largest} two-phase domain. Moreover, when we consider an \emph{intermediate}
nonhomogeneous domain between critical point and triple point, it
also seems not easy to invoke crossover arguments related to the classical
mean-field theory of critical phenomena, as we will discuss in the
final part of this paper. For example, effective values $\beta_{e}\simeq0.36-0.30$,
$\phi_{e}\simeq1.20-1.30$, and $\pi_{a,e}\simeq3.5-4$, are observed
in an extended temperature range which goes to the triple point (see
Appendix A). Such values significantly depart from mean-field ones
$\beta_{\text{MF}}=\frac{1}{2}$, $\phi_{\text{MF}}=\frac{3}{2}$,
and $\pi_{a,\text{MF}}=3$ \citet{Widom1996}. In addition to these
differences, the mean-field exponents do not satisfy hyperscaling
(i.e. explicit $d$-dependence of some scaling laws), a difficulty
precisely enhanced in the case of mean field exponents for the interfacial
properties where an explicit $\left(d-1\right)$-dependence also appears
in ''mixed'' hyperscaling laws {[}for example, see below Eq. (\ref{cross scalaw dmualpha (51)})].

Today, the estimation of parachors valid in a wide temperature range,
from a limited number of fluid-dependent parameters, still seems an
unsolved complex challenge. 

The main concern of the present paper is to clarify this situation
by only using the \emph{four} well-defined critical parameters $T_{c}$,
$p_{c}$, $Z_{c}$ and $Y_{c}$ in an asymptotic analysis of the parachor
correlations approaching the critical point. A joint objective is
to suppress asymptotical requirement for any other unknown adjustable
parameter in a well-defined extension of the VLE line close to the
critical point, substituting then the two-scale-factor universality
of dimensonless fluids to the four-parameter corresponding-states
principle to justify the observed master (i;e. unique) parachor function.

The paper is organized as follows. In Section 2, we observe, by application
of the scale dilatation method, the asymptotic master critical behaviors
for interfacial properties close to the critical point. The scale
dilatation method only uses a minimal set made of four critical scale
factors (neglecting here quantum effects in light fluids such as helium
3 \citet{Garrabos2006qe}), noted $Q_{c}^{\text{min}}=\left\{ \beta_{c}^{-1},\alpha_{c},Z_{c},Y_{c}\right\} $
and defined in the next paragraph {[}see Eqs. (\ref{energy unit (16)})
to (\ref{isochoric factor (19)})]. In Section 3, we unambiguously
separate each scale factor contribution in the estimation of either
the surface tension amplitude (only $Y_{c}$ -dependent), or the Ising-like
parachor (only $Z_{c}$-dependent). We conclude in Section 4. Appendix
A gives a complementary practical route to extend the analysis over
the complete VLE line.

\section{Master singular behaviors of interfacial properties}

\subsection{Asymptotic singular behavior of interfacial properties}

Close to the gas-liquid critical point, the asymptotic singular behavior
of thermophysical properties are generally characterized by Wegner-like
expansions \citet{Wegner1972} in terms of the following two relevant
physical variables \citet{Widom1965,Levelt1981} \begin{equation}
\Delta\tau^{*}=\frac{T-T_{c}}{T_{c}}\label{deltataustar field (2)}\end{equation}
 and\begin{equation}
\Delta\tilde{\rho}=\frac{\rho-\rho_{c}}{\rho_{c}}\label{deltatarhotilde OP (3)}\end{equation}
where subscript $c$ refers to a critical property. $\Delta\tau^{*}$
and $\Delta\tilde{\rho}$ are the temperature field and the order
parameter density, respectively, of the liquid-gas transition. Therefore,
in the non-homogeneous phase ($\Delta\tau^{*}<0$), along the critical
isochore ($\rho=\rho_{c}$), the asymptotic singular behaviors of
the symmetrized order parameter density $\Delta\rho_{LV}$ and the
interfacial tension $\sigma$ read as follows \begin{equation}
\Delta\rho_{LV}=2\rho_{c}B\left|\Delta\tau^{*}\right|^{\beta}\left[1+\overset{i=\infty}{\underset{i=1}{\sum}}B_{i}\left|\Delta\tau^{*}\right|^{i\Delta}\right],\label{Wegner eq deltarho (4)}\end{equation}
 \begin{equation}
\sigma=\sigma_{0}\left|\Delta\tau^{*}\right|^{\phi}\left[1+\overset{i=\infty}{\underset{i=1}{\sum}}\sigma_{i}\left|\Delta\tau^{*}\right|^{i\Delta}\right],\label{Wegner eq sigma (5)}\end{equation}
The Ising-like universal values of the critical exponents are $\beta\approxeq0.326$,
$\phi\approxeq1.26$, while $\Delta\approxeq0.51$ \citet{Guida1998}
is the Ising-like universal value of the lowest confluent exponent.
The amplitudes $B$, $B_{i}$, $\sigma_{0}$, and  $\sigma_{i}$,
are fluid-dependent quantities which benefit from accurate theoretical
predictions of their universal combinations and universal ratios only
valid in the Ising-like preasymtotic domain \citet{Bagnuls1985,Bagnuls1987},
where the Wegner expansion are restricted to the first order term
of the confluent corrections to scaling governed by the lowest confluent
exponent $\Delta$.

\begin{figure}
\includegraphics[width=80mm,keepaspectratio]{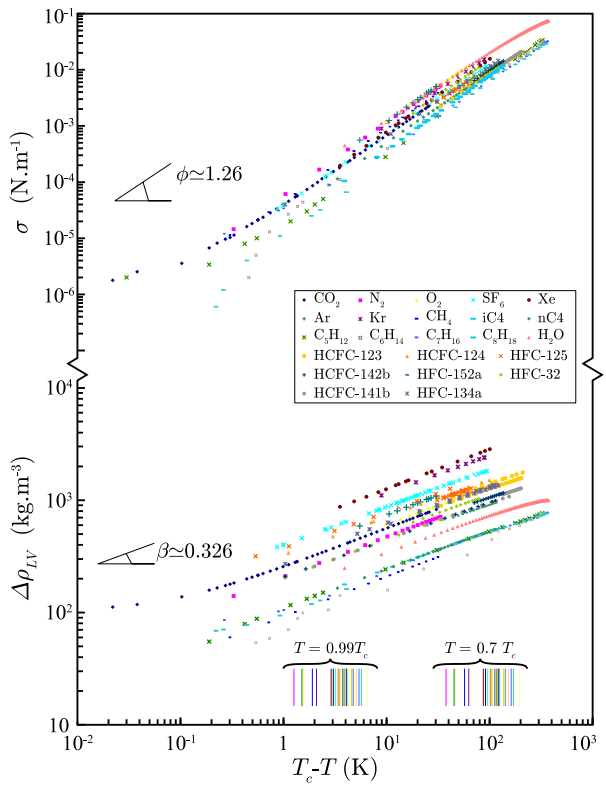}

\caption{(Color online) a) Singular behavior (log-log scale) of the interfacial
tension $\sigma$ (expressed in $\text{N}\,\text{m}^{-1}$) and b)
singular behavior of the symmetrized order parameter density $\Delta\rho_{LV}$
(expressed in $\text{kg}\,\text{m}^{-3}$), as a function of the temperature
distance $T_{c}-T$ (expressed in $\text{K}$), for one-component
fluids (see text, Appendix A and references in Table I). Inserted
Table gives colour indexation of each selected fluid. The temperature
axis is labelled by vertical arrows which indicate the practical values
$T=0.99\, T_{c}$ and $T=0.7\, T_{c}$, respectively (see text for
details). \label{fig01-parachor}}

\end{figure}

Now, we introduce the squared capillary length $\left(\ell_{Ca}\right)^{2}$
{[}also called the Sugden factor \citet{Sugden1921,Sugden1924}, noted
$S_{g}$], related to $\sigma$ and $\Delta\rho_{LV}$ by\begin{equation}
\left(\ell_{Ca}\right)^{2}\equiv S_{g}=\frac{2\sigma}{g\Delta\rho_{LV}},\label{force balance in g field (6)}\end{equation}
where $g$ is the gravitational acceleration. Equation (\ref{force balance in g field (6)})
expresses the balance between interfacial forces and volumic forces
which define the shape and position of the liquid-gas interface in
a gravity field of constant acceleration $g$. The asymptotic singular
behavior  of $S_{g}$ can be read as a Wegner-like expansion \begin{equation}
S_{g}=S_{0}\left|\Delta\tau^{*}\right|^{\varphi}\left[1+\overset{i=\infty}{\underset{i=1}{\sum}}S_{i}\left|\Delta\tau^{*}\right|^{i\Delta}\right].\label{Wegner eq sudgen (7)}\end{equation}
leading to the scaling law\begin{equation}
\varphi=\phi-\beta\label{phi vs mu beta (8)}\end{equation}
with $\varphi\approxeq0.934$, and to the canonical amplitude combination\begin{equation}
S_{0}=\frac{\sigma_{0}}{g\rho_{c}B}\label{interfacial ampli balance (9)}\end{equation}
Equations (\ref{Wegner eq deltarho (4)}) to (\ref{interfacial ampli balance (9)})
are of basic interest as well for measurement techniques of interfacial
properties as for theoretical crossover descriptions when the temperature
distance to $T_{c}$ takes a finite value (see Appendix A). 

The singular behavior of the Sugden factor $S_{g}$ as a function
of the temperature distance $T_{c}-T$ was illustrated in Fig. 1 of
Ref. \citet{Garrabos2007cal} for  about twenty pure compounds selected
among inert gases, normal compounds and highly associating polar fluids.
For the references of the $S_{g}$ and $\sigma$ data sources see
the reviews of Refs. \citet{Gielen1984,LeNeindre2002,Garrabos2007cal,Moldover1985}.
Here, we have added the data sources \citet{Okada1986,Okada1987,Higashi1992,Okada1995,Moldover1988,Chae1990}
of some hydrofluorocarbons (HFCs) and hydrocholorofluorocarbons (HCFCs)
for related discussion in Appendix A. The raw data for the surface
tension $\sigma\left(T_{c}-T\right)$ and the symmetrized order parameter
density $\Delta\rho_{LV}\left(T_{c}-T\right)$ are reported in Figs.
\ref{fig01-parachor}a and \ref{fig01-parachor}b, respectively. For
the $\Delta\rho_{LV}$ data sources see for example the references
given in Refs. \citet{Broseta2005}, \citet{LeNeindre2002}, and \citet{Okada1986,Okada1987,Higashi1992,Okada1995,Moldover1988,Chae1990}.
In each case, the universal Ising-like slope of the asymptotic singular
behavior appears compatible with the experimental results. From these
figures, it is also expected about one-decade variation for each fluid-dependent
amplitude $2\rho_{c}B\left(T_{c}\right)^{-\beta}$ and $\sigma_{0}\left(T_{c}\right)^{-\phi}$
of the leading terms of Eqs. (\ref{Wegner eq deltarho (4)}) and (\ref{Wegner eq sigma (5)}),
respectively. 

As in the Sugden factor case \citet{Garrabos2007cal}, it also appears
evident that the raw data used in present Fig. \ref{fig01-parachor}
cover a large temperature range of the coexisting liquid-vapor phases
since this range approaches the triple point temperature $T_{\text{TP}}$.
We have then adopted the same practical distinction between asymptotic
critical range and triple point region in the temperature axis, using
vertical arrows for the temperature distances where $T=0.7\, T_{c}$
(i.e. the temperature distance where the fluid-dependent acentric
factor $\omega$ is defined) and $T=0.99\, T_{c}$. In a large temperature
range defined by $0.3\leq\left|\Delta\tau^{*}\right|\leq1-\frac{T_{\text{TP}}}{T_{c}}$,
the nonuniversal nature of each fluid is certainly dominant (see for
example Appendix A), while, in the temperature range $\left|\Delta\tau^{*}\right|\leq0.01$,
the singular behavior descriptions by Wegner like expansions, and
asymptotic two-scale-factor universality of their restricted two term
form, hold.

\begin{figure}

\includegraphics[width=80mm,keepaspectratio]{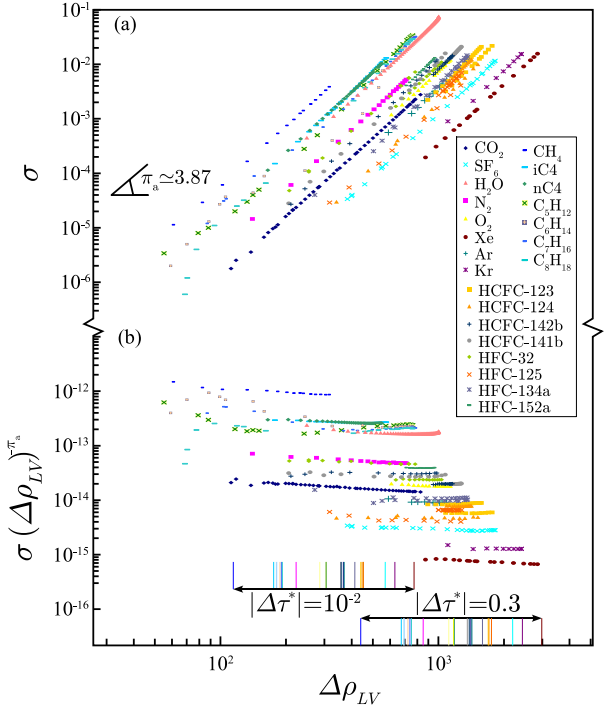}

\caption{(Color on line) a) Asymptotic singular behavior of $\sigma$ as a
function of $\Delta\rho_{LV}$ obtained from Fig. \ref{fig01-parachor}
b) As in a) for the confluent quantity $\frac{\sigma}{\left(\Delta\rho_{LV}\right)^{\pi_{a}}}$
(see caption of Fig. \ref{fig01-parachor} and text for details and
inserted Table for fluid colour indexation).\label{fig02-parachor}}

\end{figure}

Selecting $\sigma$ and $\Delta\rho_{LV}$ measurements at identical
values of $T_{c}-T$, we have constructed the corresponding $\sigma$-$\Delta\rho_{LV}$
data pairs. The singular behavior of $\sigma\left(\Delta\rho_{LV}\right)$
is illustrated in Fig. \ref{fig02-parachor}a, while the corresponding
behavior of $\sigma\left(\Delta\rho_{LV}\right)^{-\pi_{a}}$ as a
function of $\Delta\rho_{LV}$ is given in Fig. \ref{fig02-parachor}b,
as usually made to enlighten the contribution of the confluent corrections
to scaling and to have better estimation of the uncertainty attached
to the value of the leading amplitude. Simultaneously, from xenon
to n-octane, we also underline about a three-decade variation for
the fluid-dependent amplitude of this quantity, leading to about a
half-decade variation of the effective parachor (expressed in $\left[\text{J}\,\text{mole}^{-1}\,\text{m}^{3\pi_{a}-2}\right]$
unit when $\sigma$, $\Delta\rho_{LV}$, and $M_{\text{mol}}$ are
expressed in $\left[\text{J}\,\text{m}^{-2}\right]$ (or $\left[\text{N}\,\text{m}^{-1}\right]$),
$\left[\text{kg}\,\text{m}^{-3}\right]$, and $\left[\text{kg}\,\text{mole}^{-1}\right]$,
respectively). Accordingly, by straightforward elimination of $\left|\Delta\tau^{*}\right|$
in Eqs. (\ref{Wegner eq deltarho (4)}) to (\ref{Wegner eq sudgen (7)}),
we find the exact Ising-like asymptotic form\begin{equation}
\begin{array}{cl}
\sigma= & \left(\frac{P_{0}}{M_{\text{mol}}}\Delta\rho_{LV}\right)^{\pi_{a}}\\
 & \left\{ 1+P_{1}\left(\Delta\rho_{LV}\right)^{\frac{\Delta}{\beta}}+\mathcal{O}\left[\left(\Delta\rho_{LV}\right)^{\frac{2\Delta}{\beta}}\right]\right\} ,\end{array}\label{sigma vs op (10)}\end{equation}
where (as previously mentioned in the introduction part) \begin{equation}
\pi_{a}=\frac{\phi}{\beta}=\frac{\varphi}{\beta}+1\approxeq3.87,\label{pia expo vs mubeta (11)}\end{equation}
The Ising-like asymptotic value $P_{0}$ of the effective parachor
$P_{e}$ can be estimated using the leading terms of Eqs. (\ref{Wegner eq deltarho (4)})
to (\ref{Wegner eq sudgen (7)}) and reads\begin{equation}
P_{0}=\left(\sigma_{0}\right)^{\frac{\beta}{\phi}}\frac{M_{\text{mol}}}{2\rho_{c}B}=\left(gS_{0}\right)^{\frac{\beta}{\phi}}\frac{M_{\text{mol}}}{2\left(\rho_{c}B\right)^{1-\frac{\beta}{\phi}}},\label{parachor amplitude0 (12)}\end{equation}
depending on the pair of selected variables, either $\left\{ \sigma;\Delta\rho_{LV}\right\} $
or $\left\{ S_{g};\Delta\rho_{LV}\right\} $. Equations (\ref{sigma vs op (10)})
and (\ref{pia expo vs mubeta (11)}) clearly demonstrate the critical
scaling nature of Eq. (\ref{parachor correlation (1)}), with an essential
consequence: \emph{the parachor $P_{0}$ is a nonuniversal leading
amplitude which must satisfy the two-scale-factor universality of
the Ising-like universality class}. 

How to estimate the parachor appears thus as a basic question in a
sense that only two nonuniversal leading amplitudes are sufficient
to characterize the complete singular behavior of a one-component
fluid when $T\rightarrow T_{c}$ and  $\Delta\rho_{LV}\rightarrow0$.

\subsection{The basic set of fluid-dependent parameters}

\begin{table*}
\begin{tabular}{|c|c|c|c|c|c|c|c|c|c|}
\hline 
Fluid & $m_{\bar{p}}$ & $T_{c}$ & $p_{c}$ & $\rho_{c}$ & $\gamma_{c}^{'}$ & $\left(\beta_{c}\right)^{1}$ & $\alpha_{c}$ & $Z_{c}$ & $Y_{c}$\tabularnewline
 & $\left(10^{-26}\,\text{kg}\right)$ & $\left(\text{K}\right)$ & $\left(\text{MPa}\right)$ & $\left(\text{kg}\,\text{m}^{-3}\right)$ & $\left(\text{MPa}\,\text{K}^{-1}\right)$ & $\left(10^{-21}\,\text{J}\right)$ & $\left(\text{nm}\right)$ &  & \tabularnewline
\hline
\hline 
Ar & $6.6335$ & $150.725$ & $4.865$ & $535$ & $0.19025$ & $2.08099$ & $0.7535$ & $0.289871$ & $4.89423$\tabularnewline
\hline 
Kr & $13.9153$ & $209.286$ & $5.500$ & $910$ & $0.1562$ & $2.8895$ & $0.8069$ & $0.291065$ & $4.94372$\tabularnewline
\hline 
Xe & $21.8050$ & $289.740$ & $5.840$ & $1113$ & $0.1185$ & $4.0003$ & $0.8815$ & $0.286010$ & $4.87914$\tabularnewline
\hline 
N$_{2}$ & $4.6517$ & $126.200$ & $3.400$ & $314$ & $0.1645$ & $1.74258$ & $0.8002$ & $0.289078$ & $5.10585$\tabularnewline
\hline 
O$_{2}$ & $5.3136$ & $154.58$ & $5.043$ & $436$ & $0.1953$ & $2.13421$ & $0.7508$ & $0.287972$ & $4.98641$\tabularnewline
\hline 
CO$_{2}$ & $7.3080$ & $304.107$ & $7.732$ & $467.8$ & $0.173$ & $4.19907$ & $0.8289$ & $0.274352$ & $6.13653$\tabularnewline
\hline 
SF$_{6}$ & $24.2555$ & $318.687$ & $3.76$ & $741.5$ & $0.084$ & $4.40062$ & $1.054$ & $0.281243$ & $6.11960$\tabularnewline
\hline 
H$_{2}$O & $2.9969$ & $647.067$ & $22.046$ & $322.8$ & $0.2676319$ & $8.93373$ & $0.7400$ & $0.229117$ & $6.85520$\tabularnewline
\hline 
C$_{2}$H$_{4}$ & $4.658$ & $282.345$ & $5.042$ & $214.5$ & $0.11337$ & $3.89820$ & $0.91781$ & $0.28131$ & $5.34856$\tabularnewline
\hline 
CH$_{4}$ & $2.6640$ & $190.564$ & $4.59920$ & $162.7$ & $0.14746$ & $2.63102$ & $0.8301$ & $0.285752$ & $4.981927$\tabularnewline
\hline 
C$_{2}$H$_{6}$ & $4.99324$ & $305.322$ & $4.872$ & $206.58$ & $0.10304$ & $4.21554$ & $0.95290$ & $0.27935$ & $5.45505$\tabularnewline
\hline 
C$_{3}$H$_{8}$ & $7.32248$ & $369.825$ & $4.2462$ & $220$ & $0.077006$ & $5.106$ & $1.063$ & $0.27679$ & $5.70688$\tabularnewline
\hline 
n-C$_{4}$H$_{10}$ & $9.6518$ & $425.38$ & $3.809$ & $229$ & $0.0643$ & $5.87301$ & $1.155$ & $0.273352$ & $6.17774$\tabularnewline
\hline 
i-C$_{4}$H$_{10}$ & $9.6518$ & $407.85$ & $3.65$ & $225$ & $0.0643$ & $5.63102$ & $1.155$ & $0.278056$ & $6.18173$\tabularnewline
\hline 
C$_{5}$H$_{12}$ & $11.9808$ & $469.70$ & $3.3665$ & $232$ & $0.0511$ & $6.48491$ & $1.244$ & $0.270875$ & $6.12956$\tabularnewline
\hline 
C$_{6}$H$_{14}$ & $14.3100$ & $507.85$ & $3.0181$ & $234$ & $0.043658$ & $7.00666$ & $1.319$ & $0.266670$ & $6.30719$\tabularnewline
\hline 
C$_{7}$H$_{16}$ & $16.6393$ & $540.13$ & $2.727$ & $234$ & $0.038068$ & $7.45731$ & $1.398$ & $0.262180$ & $6.64356$\tabularnewline
\hline 
C$_{8}$H$_{18}$ & $18.9685$ & $568.88$ & $2.486$ & $232$ & $0.033768$ & $7.85424$ & $1.467$ & $0.258978$ & $6.82776$\tabularnewline
\hline 
HFC-32 & $8.6386$ & $351.26$ & $5.782$ & $423$ & $0.124088$ & $4.849676$ & $0.9431$ & $0.243384$ & $6.53842$\tabularnewline
\hline 
HCFC-123 & $25.3948$ & $456.82$ & $3.666$ & $554$ & $0.005711$ & $6.30764$ & $1.1982$ & $0.266439$ & $6.11647$\tabularnewline
\hline 
HCFC-124 & $22.6622$ & $395.35$ & $3.615$ & $566$ & $0.063205$ & $5.45840$ & $1.1473$ & $0.265179$ & $5.91234$\tabularnewline
\hline 
HFC-125 & $19.9301$ & $339.17$ & $3.618$ & $568$ & $0.078665$ & $4.68275$ & $1.0898$ & $0.271100$ & $6.37446$\tabularnewline
\hline 
HFC-134a & $16.9426$ & $374.30$ & $4.065$ & $512.7$ & $0.083109$ & $5.16777$ & $1.0833$ & $0.259940$ & $6.65260$\tabularnewline
\hline 
HCFC-141b & $19.42$ & $477.31$ & $4.250$ & $460$ & $0.060522$ & $6.58998$ & $1.1575$ & $0.272268$ & $5.79119$\tabularnewline
\hline 
HCFC-142b & $16.6876$ & $410.26$ & $4.041$ & $447$ & $0.071972$ & $5.66426$ & $1.1191$ & $0.266338$ & $6.30691$\tabularnewline
\hline 
HFC-152a & $10.9680$ & $386.41$ & $4.512$ & $369$ & $0.086345$ & $5.33497$ & $1.0574$ & $0.251384$ & $6.39463$\tabularnewline
\hline
\end{tabular}

\caption{Critical parameters for the selected one-components fluids.\label{tab1}}

\end{table*}

As proposed by Garrabos \citet{Garrabos1982,Garrabos1985}, a phenomenological
response to the above basic question relies on the hypothesis that
the set\begin{equation}
Q_{c,a_{\bar{p}}}^{\text{min}}=\left\{ p_{c},v_{\bar{p},c},T_{c},\gamma_{c}^{'}\right\} ,\label{critical coordinate set (13)}\end{equation}
 of four critical coordinates which localize the gas-liquid critical
point on the $p,v_{\bar{p}},T$ phase surface, contains all the needed
critical information to calculate any nonuniversal leading amplitude
of the selected fluid (here neglecting the quantum effects \citet{Garrabos2006qe}
to simplify the presentation). The mass $m_{\bar{p}}$ of each molecule
is also hypothesized known to infer the total amount $N$ of fluid
particles by measurements of the fluid total mass $M=N_{A}m_{\bar{p}}$.
$p$ ($p_{c}$) is the (critical) pressure. $v_{\bar{p}}=\frac{\bar{v}}{N_{A}}=\frac{m_{\bar{p}}}{\rho}$
($v_{\bar{p},c}=\frac{m_{\bar{p}}}{\rho_{c}}$) is the molecular volume
(critical volume). The total volume $V=Nv_{\bar{p}}$ is the extensive
variable conjugated to $p$. $\gamma_{c}^{'}=\left[\left(\frac{\partial p}{\partial T}\right)_{v_{\bar{p},c}}\right]_{\text{CP}}$
is the common critical direction in the $p;T$ diagram of the critical
isochore and the saturation pressure curve $p_{\text{sat}}\left(T\right)$
at critical point (CP), thus defined by\begin{equation}
\gamma_{c}^{'}=\left(\frac{\partial p}{\partial T}\right)_{\rho_{c},T\rightarrow T_{c}^{+}}=\left(\frac{dp_{\text{sat}}}{dT}\right)_{T\rightarrow T_{c}^{-}}\label{isochoric critical direction (14)}\end{equation}
Rewriting Eq. (\ref{critical coordinate set (13)}) as a four-scale-factor
set \begin{equation}
Q_{c}^{\text{min}}=\left\{ \beta_{c}^{-1},\alpha_{c},Z_{c},Y_{c}\right\} \label{minimal critical set (15)}\end{equation}
where,\begin{equation}
\beta_{c}^{-1}=k_{B}T_{c}\sim\left[\text{energy}\right]\label{energy unit (16)}\end{equation}
\begin{equation}
\alpha_{c}=\left(\frac{k_{B}T_{c}}{p_{c}}\right)^{\frac{1}{d}}\sim\left[\text{length}\right]\label{length unit (17)}\end{equation}
\begin{equation}
Z_{c}=\frac{p_{c}m_{\bar{p}}}{k_{B}T_{c}\rho_{c}}\label{compression factor (18)}\end{equation}
 \begin{equation}
Y_{c}=\gamma_{c}^{'}\frac{T_{c}}{p_{c}}-1\label{isochoric factor (19)}\end{equation}
 we introduce the energy unit {[}$\beta_{c}^{-1}$], the length unit
{[}$\alpha_{c}$], the (isothermal) scale factor {[}$Z_{c}$] of the
order parameter density {[}see Eq. (\ref{deltatarhotilde OP (3)})]
along the critical isothermal line, and the (isochoric) scale factor
{[}$Y_{c}$] of the temperature field {[}see Eq. (\ref{deltataustar field (2)})]
along the critical isochoric line.

Table \ref{tab1} provides values of the critical parameters involved
in Eqs. (\ref{critical coordinate set (13)}) and (\ref{minimal critical set (15)})
for the $23$ pure fluids selected in this paper.

\begin{figure}
\includegraphics[width=80mm,height=80mm,keepaspectratio]{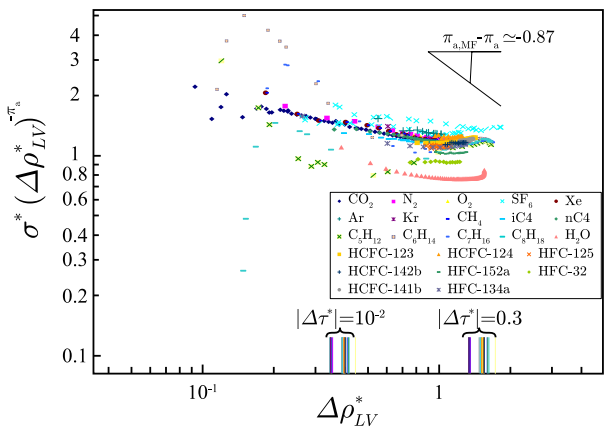}

\caption{(Color online) Singular behavior (log-log scale) of the dimensionless
quantity $\frac{\sigma^{*}}{\left(\Delta\tilde{\rho}_{LV}\right)^{\pi_{a}}}$
as a function of the symmetrized order parameter density $\Delta\tilde{\rho}_{LV}$;
inserted slope: direction difference $\pi_{a,\text{MF}}-\pi_{a}=-0.87$
from $\pi_{a,\text{MF}}=3$ (see text); arrow sets in lower horizontal
axis: $\left|\Delta\tau^{*}\right|=10^{-2}$ and $\left|\Delta\tau^{*}\right|=0.3$
(see text and caption of \ref{fig01-parachor}); Inserted Table gives
the fluid colour indexation.\label{fig03-parachor}}

\end{figure}

Indeed, our dimensional scale units of energy $\left(\beta_{c}\right)^{-1}$
{[}Eq. (\ref{energy unit (16)})], and length $\alpha_{c}$ {[}Eq.
(\ref{length unit (17)})], provide equivalent description as $T_{c}$
and $p_{c}$ of the basic (2-parameter) corresponding-states principle.
The customary dimensionless forms of $\sigma$ and $\Delta\rho_{LV}$
are $\sigma^{*}=\left(\alpha_{c}\right)^{d-1}\beta_{c}\sigma$ and
$\Delta\tilde{\rho}_{LV}=\frac{\Delta\rho_{LV}}{2\rho_{c}}$. As for
the Sugden factor case, Fig. \ref{fig03-parachor} gives distinct
curves of $\sigma^{*}\left(\Delta\tilde{\rho}_{LV}\right)^{-\pi_{a}}$
as a function of $\Delta\tilde{\rho}_{LV}$, which confirms the failure
of any description based on the two-parameter corresponding-states
principle. Moreover, the direction difference $\pi_{a,\text{MF}}-\pi_{a}\approxeq-0.87$
with a \emph{classical} power law of mean field exponent $\pi_{a,\text{MF}}=3$
also disagrees with experimental trends at large temperature distance,
as illustrated in Fig. \ref{fig03-parachor}.

On the other hand, the set $Q_{c}^{min}$ {[}Eqs. (\ref{minimal critical set (15)})
to (\ref{isochoric factor (19)})] conforms to the general description
provided from appropriate 4-parameter corresponding-states modelling,
as mentioned in our introductive part. For example, we retrieve that
the scale factor $Y_{c}$ {[}Eq. (\ref{isochoric factor (19)})] is
related to the Riedel factor $\alpha_{c,R}=\gamma_{c}^{'}\frac{T_{c}}{p_{c}}$
by $Y_{c}+1=\alpha_{c,R}$, as mentioned in our introductive part.
However, as we will extensively show in this paper, the scale dilatation
approach brings a theoretical justification to these critical parameters,
initially introduced only to build extended corresponding-states principle.

First, the microscopic meaning of $\left(\beta_{c}\right)^{-1}$ and
$\alpha_{c}$, related to the minimum value {[}$\epsilon$$\sim$$\left(\beta_{c}\right)^{-1}$],
of the interaction energy between particle pairs at equilibrium position
{[}$r_{e}$$\sim$$\frac{1}{2}$$\alpha_{c}$] , takes primary importance.
Thus, $\alpha_{c}$ appears as the mean value of the finite range
of the attractive interaction forces between particles. So that,\begin{equation}
v_{c,I}=\left(\alpha_{c}\right)^{d}\label{CIC volume (20)}\end{equation}
is the microscopic volume of the critical interaction cell at the
exact ($T=T_{c}$ and $p=p_{c}$) critical point. Using thermodynamic
properties per particle, it is straight to show that the critical
interaction cell is filled by the critical number of particles\begin{equation}
N_{c,I}=\frac{1}{Z_{c}}\label{CIC particle number (21)}\end{equation}
 since, rewriting Eq. (\ref{compression factor (18)}), we have\begin{equation}
\frac{k_{B}T_{c}}{p_{c}}=v_{c,I}=N_{c,I}\times v_{\bar{p},c}=\frac{1}{Z_{c}}\times\frac{m_{\bar{p}}}{\rho_{c}}\label{CIC law (22)}\end{equation}
Second, since in the critical phenomena description only one length
scale unit is needed to express correctly the non-trivial length dimensions
of thermodynamic and correlation variables \citet{Privman1991}, the
above microscopic analysis is of primary importance. By chosing $\alpha_{c}$,
the two dimensionless scale factors $Z_{c}$ and $Y_{c}$ are then
characteristic properties of the critical interaction cell of each
one-component fluid. Specially, $\frac{1}{Z_{c}}$ takes similar microscopic
nature of the coordination number in the lattice description of the
three-dimensional Ising systems, while $\alpha_{c}$ takes similar
microscopic nature of their lattice spacing $a_{\text{Ising}}$.

On the basis of this microscopic understanding, we are now in position
to estimate the Ising-like parachor from $Q_{c}^{\text{min}}$, using
scale dilatation of the physical fields \citet{Garrabos1982}.

\begin{table*}
\begin{tabular}{|c||c|c|c|c|c|c|c|}
\hline 
Fluid & $\phi_{e}$ & $\sigma_{0,e}$ & Ref. & $\sigma_{0}$ & $\mathcal{Z}_{\sigma,\text{exp}}$ & $\delta\mathcal{Z}_{\sigma,\text{exp}}$ & Ref\tabularnewline
 &  & $\left(10^{-3}\,\text{N}\,\text{m}^{-1}\right)$ &  & $\left(10^{-3}\,\text{N}\,\text{m}^{-1}\right)$ &  & $\%$ & \tabularnewline
\hline
\hline 
Ar &  &  &  & $31.55$ & $1.162$ & $0.5$ & \citet{Moldover1988}\tabularnewline
 &  &  &  & $30.48$ & $1.123$ & $-2.9$ & \citet{Gielen1984}\tabularnewline
 & $1.281$ & $38.07$ & \citet{Stansfield1958} & $31.42$ & $1.158$ & $0.1$ & This work\tabularnewline
 & $1.277$ & $37.78$ & \citet{Sprow1966} & $31.76$ & $1.170$ & $1.2$ & This work\tabularnewline
\hline 
Xe &  &  &  & $45.69$ & $1.179$ & $2.0$ & \citet{Moldover1985}\tabularnewline
 &  &  &  & $45.61$ & $1.177$ & $1.8$ & \citet{Gielen1984}\tabularnewline
 & $1.302\left(\pm0.006\right)$ & $62.9\left(\pm1.8\right)$ & \citet{Zollweg1971} & $47.1$ & $1.215$ & $5.1$ & This work\tabularnewline
 & $1.290$ & $53.9$ & \citet{Smith1967} (from \citet{Leadbetter1965}) & $42.68$ & $1.101$ & $-4.8$ & This work\tabularnewline
 & $1.287\left(\pm0.017\right)$ & $54.6\left(\pm0.1\right)$ & \citet{Smith1967} & $43.83$ & $1.131$ & $-2.2$ & This work\tabularnewline
\hline 
N$_{2}$ &  &  &  & $25.75$ & $1.212$ & $4.8$ & \citet{Gielen1984}\tabularnewline
\hline 
O$_{2}$ &  &  &  & $34.04$ & $1.186$ & $2.6$ & \citet{Gielen1984}\tabularnewline
\hline 
CO$_{2}$ &  &  &  & $70.05$ & $1.209$ & $4.6$ & \citet{Moldover1985}\tabularnewline
 &  &  &  & $65.93$ & $1.138$ & $-1.5$ & \citet{Gielen1984}\tabularnewline
 & $1.26$ & $76$ & \citet{Gielen1984} (from \citet{Herpin1973}) & $69.1$ & $1.193$ & $3.2$ & This work\tabularnewline
 & $1.281$ & $84.72$ & \citet{Grigull1969} & $69.92$ & $1.207$ & $4.4$ & \tabularnewline
\hline 
SF$_{6}$ &  &  &  & $47.85$ & $1.230$ & $6.4$ & \citet{Moldover1985}\tabularnewline
 &  &  &  & $44.09$ & $1.134$ & $-1.9$ & \citet{Gielen1984}\tabularnewline
 & $1.285\left(\pm0;016\right)$ & $55.13\left(\pm2.6\right)$ & \citet{Wu1973} & $46$ & $1.183$ & $2.3$ & \citet{Gielen1984}\tabularnewline
 & $1.285\left(\pm0;016\right)$ & $55.13\left(\pm2.6\right)$ & \citet{Wu1973} & $44.64$ & $1.148$ & $-0.7$ & This work\tabularnewline
 & $1.286$ & $54.28$ & \citet{Rathjen1980} & $43.78$ & $1.126$ & $-2.6$ & This work\tabularnewline
\hline 
CBrF$_{3}$ & $1.279$ & $54.05$ & \citet{Rathjen1980} & $45.02$ & $1.106$ & $-4.3$ & \tabularnewline
\hline 
CClF$_{3}$ & $1.30$ & $58.84$ & \citet{Grigull1969} & $44.5$ & $1.135$ & $-1.8$ & This work\tabularnewline
 & $1.283$ & $52.53$ & \citet{Rathjen1980} & $42.96$ & $1.096$ & $-5.2$ & This work\tabularnewline
\hline 
CHClF$_{2}$ & $1.283$ & $69.03$ & \citet{Rathjen1980} & $56.44$ & $1.139$ & $-1.5$ & This work\tabularnewline
\hline 
CCl$_{2}$F$_{2}$ & $1.283$ & $59.63$ & \citet{Rathjen1980} & $48.76$ & $1.172$ & $1.4$ & This work\tabularnewline
\hline 
CCl$_{3}$F & $1.263$ & $63.24$ & \citet{Rathjen1980} & $56.78$ & $1.196$ & $3.5$ & This work\tabularnewline
\hline 
H$_{2}$O &  &  &  & $218$ & $1.135$ & $-1.8$ & \citet{Moldover1985}\tabularnewline
 &  &  &  & $220.7$ & $1.149$ & $-0.6$ & \citet{Gielen1984}\tabularnewline
\hline 
CH$_{4}$ &  &  &  & $30.22$ & $1.012$ & $-12.4$ & \citet{Gielen1984}\tabularnewline
\hline 
C$_{2}$H$_{4}$ &  &  &  & $45.08$ & $1.166$ & $0.8$ & \citet{Moldover1985}\tabularnewline
\hline 
C$_{2}$H$_{6}$ &  &  &  & $45.15$ & $1.145$ & $-1.0$ & \citet{Moldover1985}\tabularnewline
\hline 
i-C$_{4}$H$_{10}$ &  &  &  & $45.9$ & $1.159$ & $0.2$ & \citet{Moldover1988}\tabularnewline
\hline 
$mean$ &  &  &  &  & $1.167$ & $1.0$ & \tabularnewline
$standard$ &  &  &  &  & $0.062$ & $5.4$ & \tabularnewline
\hline
\end{tabular}

\caption{Effective values of the critical exponent (column 2) and asymptotic
amplitudes of interfacial tension (column 3) from references given
in column 4. Related estimations of the leading asymptotic amplitude
$\sigma_{0}$ (column 5) (see text for details). Calculated values
of the corresponding master amplitudes $\mathcal{Z}_{\sigma,\text{exp}}=\mathcal{\sigma}_{0}\left(\alpha_{c}\right)^{2}\beta_{c}\left(Y_{c}\right)^{-\phi}$
(column 6) of the master interfacial tension (see Table \ref{tab1}
for the values of $\alpha_{c}$, $\beta_{c}$, and $Y_{c}$). The
\% differences $\delta\mathcal{Z}_{\sigma,\text{exp}}=100\left(\frac{\mathcal{Z}_{\sigma,\text{exp}}}{\mathcal{Z}_{\sigma}}-1\right)$
from the values $\mathcal{Z}_{\sigma}=1.156$ estimated from universal
amplitude combinations are given in column 7. For references see column
8. \label{tab2}}

\end{table*}

\subsection{The scale dilatation of the physical variables }

The asymptotic master critical behavior for interfacial properties
when $T\rightarrow T_{c}$ and  $\Delta\rho_{LV}\rightarrow0$, can
be observed by using the following dimensionless physical quantities\begin{equation}
\Delta\tau^{*}=k_{B}\beta_{c}\left(T_{c}-T\right)\label{temperature field (23)}\end{equation}
\begin{equation}
\Delta\mu_{\bar{p}}^{*}=\beta_{c}\left(\mu_{\bar{p}}-\mu_{\bar{p},c}\right)\label{ordering field (24)}\end{equation}
\begin{equation}
\Delta m^{*}=\left(\alpha_{c}\right)^{d}\left(n-n_{c}\right)=\left(Z_{c}\right)^{-1}\Delta\tilde{\rho}\label{order parameter (25)}\end{equation}
 and the following \emph{master} (rescaled) quantities, \begin{equation}
\mathcal{T}^{*}=Y_{c}\left|\Delta\tau^{*}\right|\label{master thermal field (26)}\end{equation}
\begin{equation}
\mathcal{H}^{*}=\left(Z_{c}\right)^{-\frac{d}{2}}\Delta\mu_{\bar{p}}^{*}=\left(Z_{c}\right)^{-\frac{1}{2}}\Delta\tilde{\mu}_{\rho}\label{master ordering field (27)}\end{equation}
\begin{equation}
\mathcal{M}^{*}=\left(Z_{c}\right)^{\frac{d}{2}}\Delta m^{*}=\left(Z_{c}\right)^{\frac{1}{2}}\Delta\tilde{\rho}\label{master order parameter (28)}\end{equation}
\begin{equation}
\Sigma^{*}\equiv\sigma^{*}\label{master surface tension (29)}\end{equation}
\begin{equation}
\mathcal{S}_{g^{*}}^{*}=g^{*}\left(Z_{c}\right)^{-\frac{3}{2}}\left(\ell_{Ca}^{*}\right)^{d-1}\label{master sugden factor (30)}\end{equation}
where, in Eqs. (\ref{master thermal field (26)}) to (\ref{master sugden factor (30)}),
we have only used the two scale factors $Y_{c}$ and $Z_{c}$ to rescale
the dimensionless quantities, with $\sigma^{*}=\left(\alpha_{c}\right)^{d-1}\beta_{c}\sigma$;
$\ell_{Ca}^{*}=\left(\alpha_{c}\right)^{-1}\ell_{Ca}$ and $g^{*}=m_{\bar{p}}\beta_{c}\alpha_{c}g$.
$\mu_{\bar{p}}$ ($\mu_{\bar{p},c}$) is the molecular chemical potential
(critical molecular chemical potential). $n$ ($n_{c}$) is the number
density (critical number density). The normalized variable $n=\frac{N}{V}$,
where $\mu_{\bar{p}}$ is conjugated to the total amount of matter
$N$, is then related to the order parameter (number) density expressing
thermodynamic properties per molecule. As mentionned in the introduction,
the (mass) density $\rho=\frac{M}{V}$, where $\mu_{\rho}$ is conjugated
to the total mass of matter $M$, is related to the order parameter
density, but expressing thermodynamic properties per volume unit.
Here we have noted $\mu_{\rho}=\frac{\mu_{\bar{p}}}{m_{\bar{p}}}$
the chemical potential per mass unit. The dimensionless form of $\mu_{\rho}$
is $\tilde{\mu}_{\rho}=\mu_{\rho}\frac{\rho_{c}}{p_{c}}$, while the
one of $\mu_{\bar{p}}$ is $\mu_{\bar{p}}^{*}=\mu_{\bar{p}}\beta_{c}$,
with $\frac{1}{Z_{c}}\mu_{\bar{p}}^{*}=\tilde{\mu}_{\rho}$. $g$
is the gravitational acceleration needed to perform conventional measurements
by capillary rise or drop techniques \citet{Sugden1924,Rowlinson 1984}.
$g^{*}$ is the dimensionless gravitational acceleration, where we
have used $\alpha_{c}\left(m_{\bar{p}}\beta_{c}\right)^{\frac{1}{2}}\sim\left[\frac{\text{mass}\left(\text{length}\right)^{2}}{\text{energy}}\right]^{\frac{1}{2}}$
as a time unit.

In Eqs. (\ref{master thermal field (26)}) and (\ref{master ordering field (27)}),
$Y_{c}$ and $Z_{c}$ are two independent scale factors that dilate
the \emph{temperature field} along the critical isochore and the \emph{ordering
field} along the critical isotherm, respectively. These Eqs. (\ref{master thermal field (26)})
and (\ref{master ordering field (27)}) are formally analogous to
analytical relations \citet{Wilson1971} linking two relevant fields
of the so-called $\Phi_{d=3}^{4}\left(n=1\right)$ model with two
physical variables of a real system belonging to the universality
class. That implicitely imposes that only a single microscopic length
is characteristic of the system \citet{Privman1991}, which then can
be related to the inverse coupling constant of the model taking appropriate
length dimension, precisely for the $d=3$ case. When a single length
is the common unit to the thermodynamic and correlation functions,
the singular part of free energy of any system belonging to the universality
class remains proportional to a universal quantity which generally
refers to the (physical) value of the critical temperature. In the
case of one-component fluids, the single characteristic length originates
from thermodynamic considerations {[}see above $\alpha_{c}$ of Eq.
(\ref{length unit (17)})]. Accordingly, the universal singular free
energy density is expressed in units of $\left(\beta_{c}\right)^{-1}=k_{B}T_{c}$
{[}see Eq. (\ref{energy unit (16)})]. The master fields $\mathcal{T}^{*}$
and $\mathcal{H}^{*}$ have similar Ising-like nature to the renormalized
fields $t$ and $h$ in field theory applied to critical phenomena.

\subsection{Master crossover behavior for interfacial properties of the one-component
fluid subclass}

The observation of the critical crossover behavior of a master property
$\mathcal{P}^{*}$ in a $\mathcal{P}^{*}-\mathcal{T}^{*}$ diagram,
generates a single curve which can be described by a master Wegner-like
expansion $\mathcal{P}^{*}\left(\mathcal{T}^{*}\right)$. Asymptotically,
i.e. for $\mathcal{T}^{*}\rightarrow0$, the universal features of
fluid singular behaviors are at least valid in the Ising-like preasymptotic
domain \citet{Bagnuls1985,Garrabos2006gb} where only two asymptotic
amplitudes and one first confluent amplitude characterize each one-component
fluid. In our present formulation of the liquid-vapor interfacial
properties, we define $\mathcal{M}_{LV}^{*}=\left(Z_{c}\right)^{\frac{3}{2}}\Delta m_{LV}^{*}=\left(Z_{c}\right)^{\frac{1}{2}}\Delta\tilde{\rho}_{LV}$,
$\Delta m_{LV}^{*}=\left(\alpha_{c}\right)^{d}\left(n_{L}-n_{V}\right)=\left(Z_{c}\right)^{-1}\Delta\tilde{\rho}_{LV}$,
$\Delta\tilde{\rho}_{LV}=\frac{\rho_{L}-\rho_{V}}{2\rho_{c}}$, and
we introduce the following master equations of interest, here restricted
to the first order of the critical confluent correction to scaling
to be in conformity with the above universal faetures,\begin{equation}
\mathcal{M}_{LV}^{*}=\mathcal{Z}_{M}\left(\mathcal{T}^{*}\right)^{\beta}\left[1+\mathcal{Z}_{M}^{\left(1\right)}\left(\mathcal{T}^{*}\right)^{\Delta}+...\right]\label{Wegner eq mop (31)}\end{equation}
 \begin{equation}
\Sigma^{*}=\mathcal{Z}_{\sigma}\left(\mathcal{T}^{*}\right)^{\phi}\left[1+\mathcal{Z}_{\sigma}^{\left(1\right)}\left(\mathcal{T}^{*}\right)^{\Delta}+...\right]\label{Wegner eq msigma (32)}\end{equation}
\begin{equation}
\mathcal{S}_{g^{*}}^{*}=\mathcal{Z}_{S}\left(\mathcal{T}^{*}\right)^{\varphi}\left[1+\mathcal{Z}_{S}^{\left(1\right)}\left(\mathcal{T}^{*}\right)^{\Delta}+...\right]\label{Wegner eq msugden (33)}\end{equation}
with their interrelation {[}see Eqs. (\ref{phi vs mu beta (8)}) and
(\ref{force balance in g field (6)})]\begin{equation}
\mathcal{S}_{g^{*}}^{*}=\frac{\Sigma^{*}}{\mathcal{M}_{LV}^{*}}\label{msugden balance (34)}\end{equation}

The master asymptotic behaviors of $\mathcal{M}_{LV}^{*}$ and $\mathcal{S}_{g^{*}}^{*}$
as a function of $\mathcal{T}^{*}$ were observed and analyzed in
Refs. \citet{Garrabos2002} and \citet{Garrabos2007cal} for several
pure fluids. The corresponding master amplitudes take the values $\mathcal{Z}_{M}\approxeq0.468\,\left(\pm0.001\right)$
and $\mathcal{Z}_{S}\approxeq2.47\,\left(\pm0.17\right)$ (for the
quoted error-bars see Refs. \citet{Garrabos2002,Garrabos2007cal}).
>From Eq. (\ref{msugden balance (34)}), the leading master amplitude
for the surface tension case is then $\mathcal{\mathcal{Z}_{\sigma}}=\mathcal{Z}_{M}\mathcal{Z}_{S}\approxeq1.156\,\left(\pm0.087\right)$.
Here, using a method similar to the one applied to the Sugden factor
case, we show in Table \ref{tab2} that this master value is compatible
with the results obtained from experiments performed sufficiently
close to the critical point. Therefore, only surface tension measurements
already analyzed by Moldover \citet{Moldover1985} and Gielen et al
\citet{Gielen1984} are considered. The respective effective values
$\phi_{e}$ and $\mathcal{\sigma}_{0,e}$ of the exponent-amplitude
pair corresponding to data fits by an effective power law $\sigma=\sigma_{0,e}\left|\Delta\tau^{*}\right|^{\phi_{e}}$
are reported in columns 2 and 3, when needed for the present analysis
(see the corresponding Refs. \citet{Stansfield1958,Leadbetter1965,Sprow1966,Smith1967,Zollweg1971,Herpin1973,Wu1973,Grigull1969,Rathjen1980}
in column 4). The estimated values of the leading amplitude $\sigma_{0}$
for the Ising value $\phi\approxeq1.260$ of the critical exponent
are given in column 5. The references reported in column 8 precise
the origin of these estimations, which are mainly dependent on the
accuracy of the interfacial property measurements in the vicinity
of $\left|\Delta\tau^{*}\right|\simeq0.01$. For example, in the present
work we have estimated $\sigma_{0}$ by the following relation $\sigma_{0}=\frac{\mathcal{\sigma}_{0,e}\left(0.01\right)^{\phi_{e}-1.260}}{1.1}$,
using data sources of columns 2 and 3. Our estimation is then compatible
with surface tension measurements at $\left|\Delta\tau^{*}\right|=0.01$,
neglecting confluent corrections in Sugden factor measurements \citet{Garrabos2007cal},
and averaging (for all the selected fluids) the confluent correction
contributions in density measurements to $10\%$ at this finite distance
to the critical temperature \citet{Garrabos2002}. The values of $\mathcal{Z}_{\sigma,\text{exp}}=\mathcal{\sigma}_{0}\left(\alpha_{c}\right)^{2}\beta_{c}\left(Y_{c}\right)^{-\phi}$
(column 6) calculated from these $\mathcal{\sigma}_{0}$ estimations
are in close agreement with our master value $\mathcal{Z}_{\sigma}=1.156$.
The residuals $\delta\mathcal{Z}_{\sigma,\text{exp}}=100\left(\frac{\mathcal{Z}_{\sigma,\text{exp}}}{\mathcal{Z}_{\sigma}}-1\right)$
(column 7) expressed in \%, are of the same order of magnitude than
the experimental uncertainties {[}see for example Refs. \citet{Moldover1985}
and \citet{Gielen1984}]. We note that the $+1\%$ residuals between
the experimental mean value $\left\langle \mathcal{Z}_{\sigma,\text{exp}}\right\rangle =1.167$
and the estimated master one $\mathcal{Z}_{\sigma}=1.156$, have a
standard deviation ($\pm5.4\%$) comparable to the experimental uncertainties
($\sim7.5\%$). However, this good agreement on the central value
is noticeable in regards to significant contributions of confluent
corrections, as reflected by the effective exponent values $\phi_{e}\approx1.28-1.31$
larger than $\phi\approxeq1.260$ at finite distance from $T_{c}$
(see also a complementary discussion related to the analysis of data
at large distance from $T_{c}$ given in Appendix A).

Similarly, each confluent amplitude $\mathcal{Z}_{M}^{\left(1\right)}$,
$\mathcal{Z}_{\sigma}^{\left(1\right)}$, and $\mathcal{Z}_{S}^{\left(1\right)}$,
takes a master constant value for all one-component fluids. Among
all $\mathcal{Z}_{P}^{\left(1\right)}$, only one is independent and
characteristic of the pure fluid subclass.

\begin{figure*}
\includegraphics[width=0.6\paperwidth,keepaspectratio]{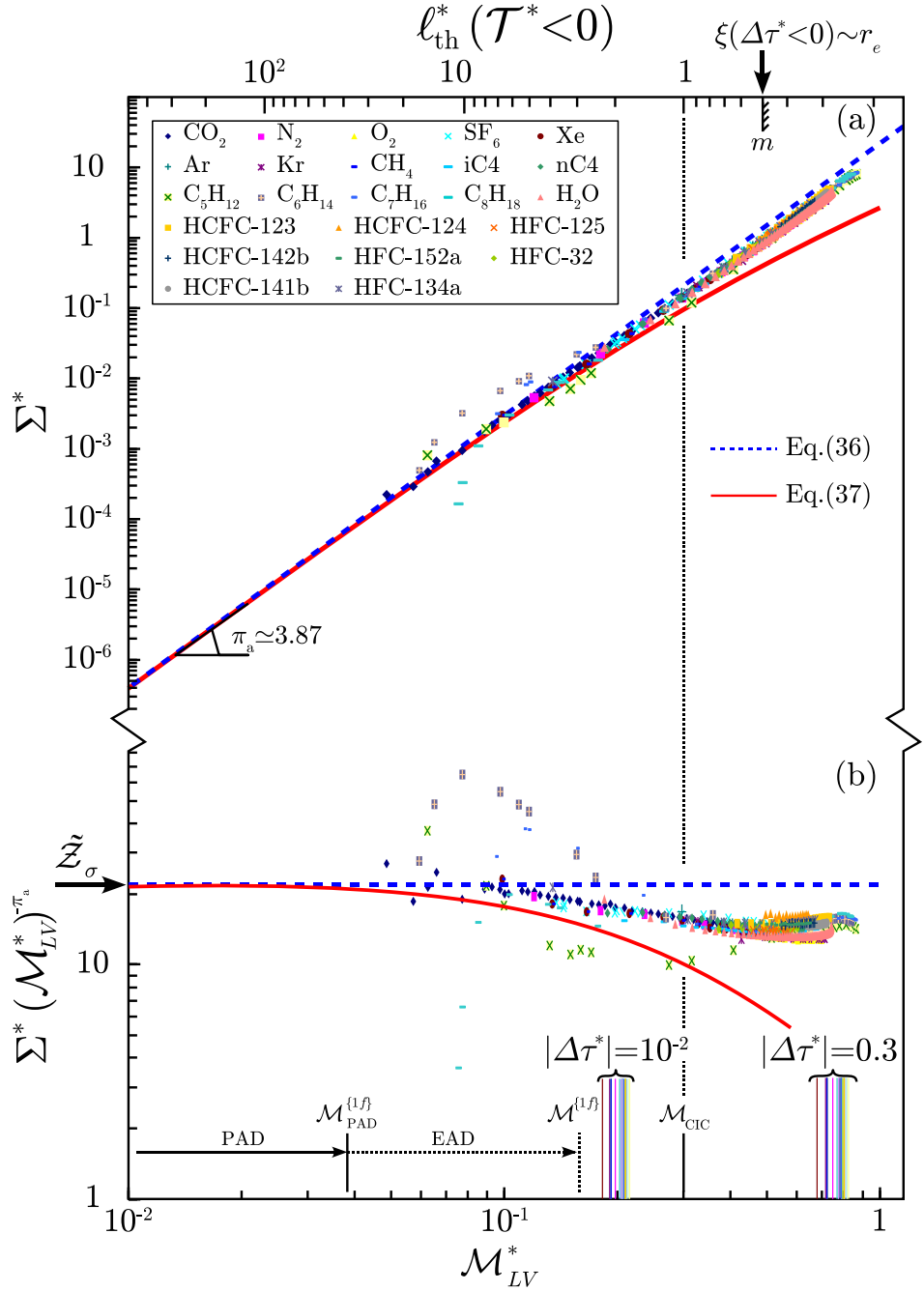}

\caption{(Colour online) a) Master singular behavior (log-log scale) of the
renormalized surface tension $\Sigma^{*}=\sigma^{*}$ {[}Eq. (\ref{master surface tension (29)})],
as a function of the renormalized (symmetrized) order parameter density
$\mathcal{M}_{LV}^{*}$ in the nonhomogeneous domain {[}see Eq. (\ref{master order parameter (28)})
and text]; b) As in a) for the {}``confluent'' quantity $\frac{\Sigma^{*}}{\left(\mathcal{M}_{LV}^{*}\right)^{\pi_{a}}}$
(see text). In a) and b): Dashed blue curve: Eq. (\ref{MastersigmaOP (36)});
full red curve: (\ref{sigmastar vs squarellestar (37)}); full arrow
with label PAD:$\mathcal{M}_{LV}^{*}<\mathcal{M}_{\text{PAD}}^{\left\{ 1f\right\} }$
extension of the preasymptotic domain; dotted arrow with label EAD:
$\mathcal{M}_{LV}^{*}<\mathcal{M}^{\left\{ 1f\right\} }$extension
of the extended asymptotic domain; The graduation of the upper horizontal
axis gives $\ell^{*}\left(\mathcal{T}^{*}<0\right)$ calculated from
theoretical crossover (see text and Ref. \citet{Garrabos2007cal});
arrow sets in lower horizontal axis:$\left|\Delta\tau^{*}\right|=10^{-2}$
and $\left|\Delta\tau^{*}\right|=0.3$, respectively; Inserted Table
gives the fluid color indexation.\label{fig04-parachor}}

\end{figure*}

Despite their great interest for the validation of theoretical predictions,
the exact forms of Eqs. (\ref{Wegner eq mop (31)}) to (\ref{Wegner eq msugden (33)})
are not essential to understand the independent scaling role of the
two scale factors $Y_{c}$ and $Z_{c}$. Moreover, our present interest
is mainly focused on the scaling nature of $Z_{c}$. More precisely,
by a variable exchange from $\Delta\tilde{\rho}$ to $\left(Z_{c}\right)^{\frac{1}{2}}\Delta\tilde{\rho}$,
the expected master behavior must be observed when the master order
parameter density $\mathcal{M}^{*}=\left(Z_{c}\right)^{\frac{1}{2}}\Delta\tilde{\rho}$
is used as a $x$- axis, generating then each {}``collapsed'' curve
of master equation $\mathcal{P}^{*}\left(\mathcal{M}^{*}\right)$
in the Ising-like preasymptotic domain, i.e. for $\mathcal{M}^{*}\rightarrow0$.
Usually, such a master behavior of the singular fluid (bulk) properties
occurs along the critical isothermal line $T=T_{c}$, i.e., for $\mathcal{T}^{*}=\Delta\tau^{*}=0$,
which provides disappearing of the scale factor $Y_{c}$ and which
only preserves the contribution of the scale factor $Z_{c}$ in the
determination of the fluid-dependent amplitudes \citet{Garrabos1985,Garrabos1986}.
However, in the nonhomogeneous domain, the order parameter density
spontaneously takes a finite value to distinguish the two coexisting
phases in equilibrium. Thus, along the critical isochore, it is also
possible to observe the master behavior of any singular (interfacial)
property expressed as a function of the {}``symmetrized'' order
parameter density, starting with the surface tension as a typical
example. 

>From $\sigma$ and $\Delta\rho_{LV}$ at identical $T_{c}-T$, we
can construct data points of master coordinates $\Sigma^{*}=\left(\alpha_{c}\right)^{d-1}\beta_{c}\sigma$
and $\mathcal{M}_{LV}^{*}=\left(Z_{c}\right)^{\frac{1}{2}}\frac{\Delta\rho_{LV}}{2\rho_{c}}$
in the $\Sigma^{*}$-$\mathcal{M}_{LV}^{*}$ diagram. As illustrated
in Fig. \ref{fig04-parachor}a, all these data collapse to define
a master behavior of $\Sigma^{*}\left(\mathcal{M}_{LV}^{*}\right)$
for $\mathcal{M}_{LV}^{*}\rightarrow0$. This collapse is well-enlightened
in Fig. \ref{fig04-parachor}b which illustrates the corresponding
master behavior of $\Sigma^{*}\left(\mathcal{M}_{LV}^{*}\right)^{-\pi_{a}}$,
without any reference to a fitting master equation. To have a better
appreciation of the real temperature range of the VLE domain, the
values of $\left|\Delta\tau^{*}\right|=0.01$ and $\left|\Delta\tau^{*}\right|=0.3$
are also given by (fluid-dependent) arrows in the $\mathcal{M}_{LV}^{*}$
axis.

Obviously, using Wegner-like expansions to eliminate $\mathcal{T}^{*}$
between Eqs. (\ref{Wegner eq mop (31)}) and (\ref{Wegner eq msigma (32)}),
the leading power law of $\Sigma^{*}$ as a function of $\mathcal{M}_{LV}^{*}$
reads as follows\begin{equation}
\Sigma^{*}=\mathcal{\widetilde{Z}}_{\sigma}\left(\mathcal{M}_{LV}^{*}\right)^{\frac{\phi}{\beta}}\left\{ 1+\mathcal{O}\left[\left(\mathcal{M}_{LV}^{*}\right)^{\frac{\Delta}{\beta}}\right]\right\} \label{master sigma vs mop (35)}\end{equation}
where $\mathcal{\widetilde{Z}}_{\sigma}=\frac{\mathcal{Z}_{\sigma}}{\left(\mathcal{Z}_{M}\right)^{\frac{\phi}{\beta}}}\approxeq21.88$.
This asymptotical amplitude is illustrated by an arrow in the vertical
axis of Fig. \ref{fig04-parachor}b and the related horizontal dashed
(blue) line indicates clearly that the observed master behavior at
finite distance to the critical point seems in {}``asymptotical''
agreement.

In addition, the formal analogy between the basic hypotheses of the
renormalization theory and the scale dilatation method makes also
easy to probe that the effective fluid crossover behavior estimated
from the massive renormalization scheme is consistent with the asymptotical
power law of Eq. (\ref{master sigma vs mop (35)})] when $\mathcal{M}_{LV}^{*}\rightarrow0$.
As a matter of fact, along the critical isochore, the theoretical
estimations of the fluid master behaviors in a whole thermal field
range $0<\left|\mathcal{T}^{*}\right|<\infty$ can be made using the
recent modifications \citet{Garrabos2006gb,Garrabos2006mcf} of the
crossover functions calculated by Bagnuls and Bervillier \citet{Bagnuls2002}
for the classical-to-critical crossover of the Ising-like universality
class. However, as already noted in the Sugden factor case \citet{Garrabos2007cal},
the theoretical function giving the classical-to-critical crossover
of the interfacial tension is not available. Thus, to estimate $\Sigma_{\text{th}}^{*}$,
we must use two alternative routes, introducing the theoretical estimations
of either $\mathcal{M}_{\text{th}}^{*}\left(\mathcal{T}^{*}<0\right)$
or $\ell_{\text{th}}^{*}\left(\mathcal{T}^{*}<0\right)$. 

A first straightforward route consists in using as an entry quantity
the theoretical value $\mathcal{M}_{\text{th}}^{*}$ of the master
order parameter density, admitting then that $\mathcal{M}_{\text{th}}^{*}\left(\mathcal{T}<0\right)\equiv\mathcal{M}_{LV}^{*}\left(\left|\mathcal{T}\right|\right)$,
an identity already validated at least in the preasymptotic domain
$\left|\mathcal{T}\right|\lesssim\mathcal{L}_{\text{PAD}}^{\left\{ 1f\right\} }$
(see Ref. \citet{Garrabos2002}). As a result, the theoretical parachor
function reads as follows\begin{equation}
\Sigma_{\text{th,}\mathcal{M}}^{*}=\mathcal{\widetilde{Z}}_{\sigma}\left(\mathcal{M}_{\text{th}}^{*}\right)^{\frac{\phi}{\beta}}\label{MastersigmaOP (36)}\end{equation}
justifying precisely the dashed blue curves in Figs. \ref{fig04-parachor}a
and b, for the complete $\mathcal{M}_{LV}^{*}$-range. Unfortunately,
at large values of $\mathcal{T}^{*}$, the theoretical function $\mathcal{M}_{\text{th}}^{*}\left(\mathcal{T}^{*}<0\right)$
is certainly not able to reproduce the experimental behavior of $\mathcal{M}_{LV}^{*}\left(\left|\mathcal{T}\right|\right)$
and, in the absence of this dedicated analysis, we cannot indicate
easily the true extension of the $\mathcal{M}_{LV}^{*}$-range where
the identity $\mathcal{M}_{\text{th}}^{*}\left(\mathcal{T}<0\right)\equiv\mathcal{M}_{LV}^{*}\left(\left|\mathcal{T}\right|\right)$
holds. 

To pass round this difficulty, a second route combines the theoretical
functions $\ell_{\text{th}}^{*}\left(\mathcal{T}^{*}<0\right)$ and
$\mathcal{M}_{\text{th}}^{*}\left(\mathcal{T}^{*}<0\right)$ in a
whole thermal field range $0<\left|\mathcal{T}^{*}\right|<\infty$,
to infer numerically the function $\ell_{\text{th}}^{*}\left(\mathcal{M}_{\text{th}}^{*}\right)$
by exchanging the $\left|\mathcal{T}^{*}\right|$-dependence of $\ell_{\text{th}}^{*}$
by the $\mathcal{M}_{\text{th}}^{*}$-dependence of $\left|\mathcal{T}^{*}\right|$
(i.e., reversing the function $\mathcal{M}_{\text{th}}^{*}\left(\mathcal{T}^{*}<0\right)$).
Anticipating the introduction of the universal number $R_{\sigma\xi}^{-}$
recalled below, the following asymptotic scaling form of the renormalized
interfacial tension\begin{equation}
\Sigma_{\text{th},\ell}^{*}=R_{\sigma\xi}^{-}\left[\ell_{\text{th}}^{*}\left(\mathcal{M}_{th}^{*}\right)\right]^{1-d}\label{sigmastar vs squarellestar (37)}\end{equation}
provides the correct asymptotic behavior of $\ell_{\text{th}}^{*}\left(\mathcal{M}_{\text{th}}^{*}\right)$
in the Ising limit $\mathcal{M}_{\text{th}}^{*}\rightarrow0$, as
illustrated by the full red curves in Figs. \ref{fig04-parachor}a
and b. However, in spite of the increasing difference observed in
Fig. \ref{fig04-parachor} between these two theoretical estimations,
or between them and the observed master behavior, when $\mathcal{M}_{LV}^{*}$
increase, we can now give a well-defined estimation of the effective
extension of the VLE domain where the master crossover behavior of
$\Sigma^{*}$have physical meaning.

Indeed, in a similar way as in our previous analysis of $S_{g}$,
the universal prefactor of Eq. (\ref{sigmastar vs squarellestar (37)})
accounts for the {}``Ising-like'' behaviors of the correlation length
$\xi\left(\Delta\tau^{*}<0\right)$ and the surface tension $\sigma\left(\left|\Delta\tau^{*}\right|\right)$,
where we have used the universal ratio $\frac{\xi\left(\Delta\tau^{*}>0\right)}{\xi\left(\Delta\tau^{*}<0\right)}=1.96$
and the universal amplitude combinations given by the products of
the interfacial tension by the squared correlation length \citet{Gielen1984,Moldover1985},
i.e.,\begin{equation}
R_{\sigma\xi}^{\pm}=\lim\left\{ \beta_{c}\sigma\left(\left|\Delta\tau^{*}\right|\right)\left[\xi\left(\Delta\tau^{*}\right)\right]^{d-1}\right\} _{\Delta\tau^{*}\rightarrow0^{\pm}}\label{Rsigmaksi (38)}\end{equation}
where $R_{\sigma\xi}^{+}\approxeq0.376=\left(1.96\right)^{2}R_{\sigma\xi}^{-}$,
so that $R_{\sigma\xi}^{-}\approxeq0.979$ (for the estimated error-bars
see also Ref. \citet{Garrabos2007cal}). The superscripts $\pm$ refer
to the singular behavior of $\xi$ above ($+$) or below ($-$) $T_{c}$.
Equation (\ref{Rsigmaksi (38)}) means that the interfacial energy
of a surface area $\xi^{d-1}$ tends to a universal value {[}expressed
in units of $\left(\beta_{c}\right)^{-1}$] for any system belonging
to the Ising-like universality class (we recall that the thickness
of the interface is then of order $\xi\left(\Delta\tau^{*}<0\right)$).
In such a description, the Wegner-like expansion of the correlation
length, along the critical isochore, above and below $T_{c}$, reads
as follows\begin{equation}
\xi=\xi_{0}^{\pm}\left|\Delta\tau^{*}\right|^{-\nu}\left[1+\overset{i=\infty}{\underset{i=1}{\sum}}\xi_{i}\left|\Delta\tau^{*}\right|^{i\Delta}\right]\label{Wegner eq ksi (39)}\end{equation}
 where the fluid-dependent amplitudes $\xi_{0}^{+}$ and $\xi_{0}^{-}$
are such that $\frac{\xi_{0}^{+}}{\xi_{0}^{-}}\approxeq1.96$ (see
above), while the contribution of the confluent corrections is hypothesized
the same above and below $T_{c}$. Accordingly, the master correlation
length $\ell_{\text{th}}^{*}=\frac{\xi}{\alpha_{c}}$ (where we neglect
here the quantum effects at the microscopic length scale of the order
of $\alpha_{c}$ \citet{Garrabos2006qe}), can be estimated using
the asymptotical modifications \citet{Garrabos2006cl,Garrabos2006mcf}
of the crossover function for the correlation length in the homogeneous
domain \citet{Bagnuls2002}. This master asymptotic behavior of $\ell^{*}$
as a function of $\mathcal{T}^{*}>0$ was analyzed in Ref. \citet{Garrabos2006cl}
for seven different pure fluids, demonstrating that two specific $\ell_{\text{th}}^{*}$-values
in the range $\ell_{\text{th}}^{*}>1$ provide convenient marks to
define:

i) the extension of the Ising-like preasymptotic domain (i.e., $\ell_{\text{th}}^{*}\gtrsim80$),
where the fluid characterization is exactly conform to the universal
features calculated from the massive renormalization scheme;

ii) the extension of the effective fluid master bahavior at finite
distance to the critical temperature where $\ell_{\text{th}}^{*}\gtrsim3$.

Introducing $\ell_{\text{th}}^{*}\left(\mathcal{T}^{*}<0\right)=\frac{\ell_{\text{th}}^{*}\left(\mathcal{T}^{*}>0\right)}{1.96}$,
we can then complete the restricted two-term master forms of Eqs.
(\ref{Wegner eq mop (31)}) to (\ref{Wegner eq msugden (33)}) valid
in the Ising-like preasymptotic domain, by the following two-term
equation \begin{equation}
\ell^{*}\left(\mathcal{T}^{*}<0\right)=\mathcal{Z}_{\xi}^{-}\left|\mathcal{T}^{*}\right|^{-\nu}\left[1+\mathcal{Z}_{\xi}^{1}\left|\mathcal{T}^{*}\right|^{\Delta}+...\right]\label{Wegner eq mlcorr (40)}\end{equation}
where $\mathcal{\mathcal{Z}_{\xi}^{\mathrm{-}}}\approxeq\frac{\mathcal{Z}_{\xi}^{+}}{1.96}\approxeq0.291$
(with $\mathcal{Z}_{\xi}^{+}\approxeq0.570$) and $\mathcal{Z}_{\xi}^{1}=\mathcal{Z}_{\xi}^{1,+}=0.377$
\citet{Garrabos2006cl}. The universal features of the interfacial
properties within the Ising-like preasymptotic domain are in conformity
with the three-master amplitude characterization of the one-component
fluid subclass defined in Ref. \citet{Garrabos2006mcf}. The related
singular behaviors of the dimensionless interfacial properties of
each pure fluid can be estimated only knowing $Y_{c}$ and $Z_{c}$.
Using now the numerical function $\ell_{\text{th}}^{*}\left(\mathcal{M}_{\text{th}}^{*}\right)$,
we have also normed the upper $x$-axis in Fig. \ref{fig04-parachor}
to illustrate the singular divergence of $\ell_{\text{th}}^{*}\left(\mathcal{T}^{*}<0\right)$
in complete equivalence to the upper $x$-axis of Fig. 3 in Ref. \citet{Garrabos2007cal}.

>From $\left|\mathcal{T}^{*}\right|=\mathcal{L}_{\text{PAD}}^{\left\{ 1f\right\} }\simeq5\times10^{-4}$
where $\ell_{\text{th}}^{*}\simeq40$ {[}see Fig. 3 in Ref. \citet{Garrabos2007cal}],
the $\mathcal{M}_{LV}^{*}$-extension of the preasymptotic domain
is\begin{equation}
\mathcal{M}_{LV}^{*}<\mathcal{M}_{\text{PAD}}^{\left\{ 1f\right\} }\simeq4\times10^{-2}\label{McalPAD (41)}\end{equation}
as shown by the arrow labeled PAD in Fig. \ref{fig04-parachor}. Within
this Ising-like preasymptotic domain, the agreement between the two
theoretical estimations of $\Sigma^{*}$ are (qualitatively) conform
to the three-amplitude characterization of the universal features.
The quantitative conformity cannot be exactly accounted for within
a well-estimated theoretical erro-bar, due to the absence of theoretical
prediction for the crossover of the surface tension, large error-bar
in the estimation of the amplitude of the first-order confluent correction
term of the order-parameter density, and hypothesized contribution
of the (homogeneous) confluent corrections in the correlation length
case. Since $\alpha_{c}$ measures the shorted-range of the microscopic
molecular interaction, $\ell^{*}=\frac{\xi}{\alpha_{c}}$ gives the
relative order of magnitude of the true correlation length $\xi$,
and we can retrieve for the properties of a vapor-liquid interface
of thickness $\sim\xi$, the similar physical meaning of the two-scale
factor universality in the close vicinity of the critical point, i.e.,
when the conditions $\xi\gg\alpha_{c}$, or equivalently $\ell^{*}\gg1$,
are satisfied. Practically, the asymptotic singular behaviors of $\sigma$
(respectively $\Sigma^{*}$) and $\Delta\rho_{LV}^{*}$ (respectively
$\mathcal{M}_{LV}^{*}$), including then the first order confluent
correction to scaling as given by Eqs. (\ref{Wegner eq mop (31)})
to (\ref{Wegner eq msugden (33)}), are observed when the correlation
length in the non-homogeneous domain estimated from Eq. (\ref{Wegner eq mlcorr (40)})
is such that $\xi\gtrsim40\,\alpha_{c}$, or $\left\{ \mathcal{M}_{LV}^{*}\lesssim\mathcal{M}_{\text{PAD}}^{\left\{ 1f\right\} }\simeq0.04;\left|\mathcal{T}^{*}\right|\lesssim\mathcal{L}_{\text{PAD}}^{\left\{ 1f\right\} }\simeq0.0005\right\} $,
equivalently. This finite extension of the non-homogeneous Ising-like
preasymptotic domain, corresponds to a correlation volume $\xi^{d}$
of the fluctuating interface (of typical thickness $\sim40\, nm$,
see Table \ref{tab1}) which contains more than $6\times10^{4}$ {}``microscopic''
(i.e., $v_{c,I}$) volumes, and therefore, at least $2\times10^{5}$
cooperative particles for which the microscopic details of their molecular
interaction at the $\alpha_{c}$-scale (typically $\sim1\, nm$, see
\ref{tab1}) are then unimportant (here admitting that mean number
of fluid particles filling $v_{c,I}$ is typically $\frac{1}{Z_{c}}\simeq3.5$). 

Simarly, using $\left|\mathcal{T}^{*}\right|=\mathcal{L}^{\left\{ 1f\right\} }\simeq0.03$
{[}see Eq. (55) in Ref. \citet{Garrabos2007cal}] where $\ell_{\text{th}}^{*}\simeq3$,
the $\mathcal{M}_{LV}^{*}$-extension of the extended asymptotic domain
is\begin{equation}
\mathcal{M}_{LV}^{*}<\mathcal{M}^{\left\{ 1f\right\} }\simeq0.16\label{eq:McalEAD (42)}\end{equation}
as shown by the arrow labeled EAD in Fig. \ref{fig04-parachor}).
As expected, the master behavior of $\Sigma^{*}$ is readily observed
in this extended critical domain and in Appendix A, we give a convenient
master modification of Eq. (\ref{MastersigmaOP (36)}) to account
for it precisely. Such an extended domain of the master behavior of
the fluid subclass can be still understood since $\xi\gtrsim3\,\alpha_{c}$,
so that $\xi^{d}\gtrsim30\, v_{c,I}$, and then more than $100$ particles
in cooperative interaction. However, it is noticeable that the practical
relative values $T=0.99\, T_{c}$ of the temperature distance to $T_{c}$
frequently referred to define the critical region for each pure fluid
are not inside the effective extension of the observed master singular
behavior (see also Appendix A). 

Finally, using $\left|\mathcal{T}^{*}\right|=\mathcal{L}_{\text{CIC}}\simeq0.2$
where the size of the correlation length is equal to the size of the
critical interaction cell filled by three or four particles, i.e.,
$\ell_{\text{th}}^{*}\simeq1$ {[}see Fig. 3 in Ref. \citet{Garrabos2007cal}],
the related value of the order parameter density is\begin{equation}
\mathcal{M}_{\text{CIC}}\simeq0.3\label{McalCIC (43)}\end{equation}
as illustrated by vertical dotted line in Fig. \ref{fig04-parachor}.
We note that the pratical limit $T=0.99\, T_{c}$ with $\xi\backsim2\,\alpha_{c}$,
is in between $\mathcal{M}^{\left\{ 1f\right\} }$ and $\mathcal{M}_{\text{CIC}}$,
so that $\xi^{d}\sim8\, v_{c,I}$ only involving $\sim28$ particles
in interaction. Such a microscopic situation make questionable the
critical nature of the fluid properties measured at this finite distance
to $T_{c}$, and more generally, shows that the range $\mathcal{M}_{LV}^{*}\gtrsim0.3$
can be considered as non-Ising-like in nature. Especially, Fig. \ref{fig04-parachor}
indicates unambiguously that the temperature $T=0.7\, T_{c}$ where
the acentric factor is defined, leads to a correlation length smaller
than the mean equilibrium distance $r_{e}$ between two-interacting
particles, since the value $\ell^{*}\simeq\frac{1}{2}$ (see the hatched
limit labeled m in upper $x$-axis of Fig. \ref{fig04-parachor})
corresponds approximatively to $\xi\backsim\frac{1}{2}\alpha_{c}\backsim r_{e}$.
Appendix A provides complementary analysis of this {}``nonuniversal''
fluid crossover over the complete temperature range.

The remaining correlative problem applying the scale dilatation method
to liquid-vapor interfacial measurements, is to estimate the respective
contribution of each scale factor $Y_{c}$ and $Z_{c}$  in the fluid-dependent
amplitudes of the surface tension, expressed either as a function
of $\mathcal{T}^{*}$, or as a function of $\mathcal{M}_{LV}^{*}$.
This problem is treated in the next section.

\section{Independent scaling roles of the two scale factors}

\subsection{The thermal field dependence characterized by the $Y_c$ scale factor}

By inverting Eqs. (\ref{master thermal field (26)}) to (\ref{master sugden factor (30)}),
we can easily recover the asymptotical form for the interfacial properties
of Eqs. (\ref{Wegner eq deltarho (4)}) to (\ref{Wegner eq sudgen (7)})
from the asymptotic form of the master Eqs. (\ref{Wegner eq mop (31)})
to (\ref{Wegner eq msugden (33)}). For example, the leading physical
amplitudes $B$, $\sigma_{0}$, and $S_{0}$, can be estimated from
the following relations\begin{equation}
B=\left(Z_{c}\right)^{-\frac{1}{2}}\left(Y_{c}\right)^{\beta}\mathcal{Z}_{M}\label{B amplitude (44)}\end{equation}
 \begin{equation}
\sigma_{0}=\left(\beta_{c}\right)^{-1}\left(\alpha_{c}\right)^{1-d}\left(Y_{c}\right)^{\phi}\mathcal{Z}_{\sigma}\label{sigma0 amplitude (45)}\end{equation}
\begin{equation}
S_{0}=\left(\alpha_{c}\right)^{d-1}\left(g^{*}\right)^{-1}\left(Z_{c}\right)^{\frac{3}{2}}\left(Y_{c}\right)^{\varphi}\mathcal{Z}_{S}\label{S0 amplitude (46)}\end{equation}
Similarly, from comparison of Eqs. (\ref{Wegner eq ksi (39)}) and
(\ref{Wegner eq mlcorr (40)}), the amplitudes $\xi_{0}^{\pm}$ of
the (bulk) correlation length can be estimated from the following
relation \begin{equation}
\xi_{0}^{\pm}=\alpha_{c}\left(Y_{c}\right)^{-\nu}\mathcal{Z}_{\xi}^{\pm}\label{ksi0 amplitude (47)}\end{equation}
 As expected, the leading amplitudes are combinations of the fluid
scale factors. However, we underline the universal hyperscaling feature
of Eqs. (\ref{sigma0 amplitude (45)}) and (\ref{ksi0 amplitude (47)}),
where $\sigma_{0}$ and $\xi_{0}^{\pm}$ appear \emph{unequivocally}
related only to the scale factor $Y_{c}$. This result is obtained
from Widom's scaling law \begin{equation}
\left(d-1\right)\nu=\phi\label{Widom scaling law (48)}\end{equation}
 with $d=3$ in our present study. Using Eqs. (\ref{sigma0 amplitude (45)}),
(\ref{ksi0 amplitude (47)}) and Widom's scaling law {[}see Eq. (\ref{Widom scaling law (48)})],
it is then easy, thanks to the formal analogy between the scale dilatation
method and the analytic hypothesis of the renormalization theory,
to validate the well-known universal amplitude combination previously
introduced {[}see Eq. (\ref{Rsigmaksi (38)})]\begin{equation}
R_{\sigma\xi}^{\pm}=\beta_{c}\sigma_{0}\left(\xi_{0}^{\pm}\right)^{d-1}=\mathcal{Z}_{\sigma}\left(\mathcal{Z}_{\xi}^{\pm}\right)^{d-1}\label{universal Rsigmaksi (49)}\end{equation}

We underline also the microscopic analogy between the scale units
$\left\{ \beta_{c}^{-1},\alpha_{c}\right\} $ of the one-component
fluid and the scale units $\left\{ k_{B}T_{c},a_{\text{Ising}}\right\} $
used in Monte Carlo simulations of the simple cubic Ising-model, where
$a_{\text{Ising}}$ is the spacing lattice size \citet{Zinn1996}.
Such simulations give $\sigma_{0}=\left(\text{universal}\,\text{const}\right)\times\frac{k_{B}T_{c}}{\left(a_{\text{Ising}}\right)^{2}}$,
{[}which compares to Eq. (\ref{sigma0 amplitude (45)})], and $\xi_{0}^{+}=\left(\text{universal}\,\text{const}\right)\times a_{\text{Ising}}$,
{[}which compare to Eq. (\ref{ksi0 amplitude (47)})], to provide
a Monte Carlo estimation of the above universal ratio \citet{Privman1991,Zinn1996}.

In addition to the universal combination (\ref{universal Rsigmaksi (49)}),
we also briefly recall that equivalent universal combinations exist
between the interfacial tension amplitude $\sigma_{0}$ and the heat
capacity amplitudes $A^{\pm}$ as a following form\begin{equation}
R_{\sigma A}^{\pm}=\beta_{c}\sigma_{0}\left(A^{\pm}\right)^{\frac{d-1}{d}}\label{universal RsigmaA (50)}\end{equation}
where $R_{\sigma A}^{+}\approxeq0.275\approxeq\left(0.537\right)^{\frac{2}{3}}R_{\sigma A}^{-}$
\citet{Gielen1984,Moldover1985,Bagnuls2002}. Such universal amplitude
combinations are related to the \emph{mixed} hyperscaling laws\begin{equation}
\frac{\phi}{d-1}=\nu=\frac{2-\alpha}{d}\label{cross scalaw dmualpha (51)}\end{equation}
which give common universal features for \emph{interfacial} properties
(with dimension $d-1$) and \emph{bulk} properties (with dimension
$d=3$). $\alpha\approxeq0.11$ is the universal critical exponent
associated to the singular heat capacity. The above Eq. (\ref{cross scalaw dmualpha (51)})
can be obtained by combining Widom's scaling law,\[
\left(d-1\right)\nu=\phi\]
{[}see Eq. (\ref{Widom scaling law (48)})] and hyperscaling law\begin{equation}
d\nu=2-\alpha\label{hyperscallaw dnualpha (52)}\end{equation}
Equation (\ref{hyperscallaw dnualpha (52)}) means that the free energy
of a fluctuating bulk volume $\xi^{d}$ also tends to an universal
value {[}expressed in units of $\left(\beta_{c}\right)^{-1}$] for
any system belonging to the Ising-like universality class. We can
then focuss our interest in the singular part $\Delta c_{V,\bar{p}}\left(\Delta\tau^{*}\right)$
of the heat capacity at constant volume expressed \emph{per particle},
along the critical isochore (ignoring the classical background part
of the total heat capacity at constant volume). Indeed, the heat capacity
per particle $c_{V,\bar{p}}\sim\left[\frac{\text{particle}\,\text{energy}}{\text{temperature}\,\text{increment}}\right]$
is the unique thermodynamic property which can be made dimensionless
by only using the {}``universal'' Boltzmann factor $k_{B}$, i.e.
without reference to $\alpha_{c}$ and $\left(\beta_{c}\right)^{-1}$.
Therefore, when the singular heat capacity at constant volume, normalized
per particle, obeys the asymptotic power law\begin{equation}
\Delta c_{V,\bar{p}}=\frac{A_{0,\bar{p}}^{\pm}}{\alpha}\left|\Delta\tau^{*}\right|^{-\alpha}\left[1+\mathcal{O}\left\{ \left|\Delta\tau^{*}\right|^{\Delta}\right\} \right]\label{Wegner eq cvparticle (53)}\end{equation}
along the critical isochore, one among the two \emph{dimensionless
amplitudes} $\frac{A_{0,\bar{p}}^{+}}{k_{B}}$ and $\frac{A_{0,\bar{p}}^{-}}{k_{B}}$
\emph{is mandatorily a characteristic fluid-particle-dependent number}.
>From the basic hypothesis of the scale dilatation method, this number
should be related to $Y_{c}$ and $Z_{c}$ in a well-defined manner
to account for extensive and critical natures of the fluid system,
as will be shows below {[}see Eq. (\ref{Aparticle amplitude (56)})].

For a 3D Ising-system \citet{Privman1991}, the singular part of the
heat capacity normalized by $k_{B}$ can be expressed in units of
$\left(a_{\text{Ising}}\right)^{d}$. In the case of the one-component
fluid, the normalized heat capacity is expressed in units of $\left(\alpha_{c}\right)^{d}$,
which is the volume of the critical interaction cell. For the one-component
fluid subclass, the number of particles filling the critical interaction
cell is $\frac{1}{Z_{c}}$, leading then to define the singular part
of the master heat capacity per critical interaction cell volume (neglecting
quantum effects), as follows\begin{equation}
\mathcal{C}_{S}^{*}=\left[\left(Y_{c}\right)^{2}Z_{c}\right]^{-1}\Delta c_{V,\bar{p}}^{*}\label{dimensionless mcvs (54)}\end{equation}
 with $\Delta c_{V,\bar{p}}^{*}=\frac{\Delta c_{V,\bar{p}}}{k_{B}}$.
As needed from thermodynamics, such master heat capacity corresponds
to a second derivative, $\mathcal{C}_{S}^{*}=-\frac{\partial^{2}\mathcal{A}_{S}^{*}\left(\mathcal{T}^{*}\right)}{\partial\mathcal{T}^{*2}}$
of a master free energy $\mathcal{A}_{S}^{*}\left(\mathcal{T}^{*}\right)$
with respect to master thermal field $\mathcal{T}^{*}$ (we do not
consider here the critical contribution of an additive master constant
$\mathcal{A}_{c}^{*}$ and the regular background contribution $\mathcal{A}_{B}^{*}\left(\mathcal{T}^{*}\right)$
characteristic of each one-component fluid). Admitting now that the
leading singular part of the master free energy behaves as $\mathcal{A}_{S}^{*}=\frac{\mathcal{Z}_{A}^{\pm}}{\alpha\left(1-\alpha\right)\left(2-\alpha\right)}\left|\mathcal{T}^{*}\right|^{2-\alpha}\left[1+\mathcal{O}\left\{ \left|\mathcal{T}^{*}\right|^{\Delta}\right\} \right]$,
the master asymptotic behavior of the heat capacity reads as follows
(ignoring the critical and background contributions due to derivatives)\begin{equation}
\mathcal{C}_{S}^{*}=\frac{\mathcal{Z}_{A}^{\pm}}{\alpha}\left|\mathcal{T}^{*}\right|^{-\alpha}\left[1+\alpha\mathcal{Z}_{A}^{1,\pm}\left|\mathcal{T}^{*}\right|^{\Delta}\right]\label{Wegner eq mcvs (55)}\end{equation}
 The constant values of the master amplitudes are $\mathcal{Z}_{A}^{+}\approxeq0.1057\approxeq0.537\mathcal{Z}_{A}^{-}$,
so that $\mathcal{Z}_{A}^{-}\approxeq0.1967$, in conformity with
their universal ratio $\frac{\mathcal{Z}_{A}^{+}}{\mathcal{Z}_{A}^{-}}\approxeq0.537$
\citet{Bagnuls2002} for $d=3$. Therefore, the corresponding typical
asymptotic amplitudes $A_{0,\bar{p}}^{\pm}$ in Eq. (\ref{Wegner eq cvparticle (53)})
can be estimated from\begin{equation}
\frac{1}{Z_{c}}\frac{A_{0,\bar{p}}^{\pm}}{k_{B}}=\left(Y_{c}\right)^{2-\alpha}\mathcal{Z}_{A}^{\pm}\label{Aparticle amplitude (56)}\end{equation}
where the respective scale factor contributions (i.e., the master
nature of the critical interaction cell volume characterized by $\frac{1}{Z_{c}}$,
and the field scale dilatation along the critical isochore characterized
by $Y_{c}$), are well-identified using a single amplitude which \emph{de
facto} characterizes the particle. 

Now, for comparison with standard notations used in the literature
on fluid-related critical phenomena where all the thermodynamic potentials
are taken per unit volume, and not per particle, we also introduce
the heat capacity at constant volume per unit volume\emph{,} $\Delta c_{V=1}=\frac{\Delta c_{V,\bar{p}}}{v_{\bar{p},c}}$
(labeled here with the subscript $V=1$). Expressed in our above unit
length scale {[}Eq. (\ref{length unit (17)})], the associated dimensionless
form is\begin{equation}
\Delta c_{V=1}^{*}=\frac{\Delta c_{V,\bar{p}}}{k_{B}}\times\frac{1}{v_{\bar{p},c}\left(\alpha_{c}\right)^{-d}}=\frac{\Delta c_{V=1}}{k_{B}\left(\alpha_{c}\right)^{-d}}\label{(57)}\end{equation}
 Obviously, $\Delta c_{V=1}^{*}$ is strictly identical to the usual
dimensionless form $\Delta c_{V}^{*}=\Delta C_{V}\frac{T_{c}}{Vp_{c}}$
of the total singular heat capacity $\Delta C_{V}=N$$\Delta c_{V,\bar{p}}$
of the constant total fluid volume $V$, filled with the constant
amount $N$ of particles. Using the total Helmholtz free energy $A\left(T,V,N\right)$
where $T,V,N$ are the selected three natural variables , we obtain
$\frac{C_{V}}{T}=-\left(\frac{\partial^{2}A}{\partial T^{2}}\right)_{V,N}$
. From the corresponding quantities normalized per unit volume, $\frac{A\left(T,1,n=\frac{N}{V}\right)}{V}$,
we obtain $\frac{C_{V}}{VT}=-\left(\frac{\partial^{2}\left(\frac{A}{V}\right)}{\partial T^{2}}\right)_{n}$
, which can be considered to define their related singular dimensionless
parts $\left(\frac{A}{V}\right)_{S}^{*}=\frac{A\left(\frac{T}{T_{c}},1,n_{c}=\frac{N_{c}}{V}\right)}{Vp_{c}}$
and $\frac{T_{c}\Delta c_{V}^{*}}{T}=-\left(\frac{\partial^{2}\left(\frac{A}{V}\right)_{S}^{*}}{\partial\left(\frac{T}{T_{c}}\right)^{2}}\right)_{n=n_{c}}$
along the critical isochore. Admitting now that the leading singular
term of the free energy divergence behaves as $\left(\frac{A}{V}\right)_{S}^{*}=\frac{A^{\pm}}{\alpha\left(1-\alpha\right)\left(2-\alpha\right)}\left|\Delta\tau^{*}\right|^{2-\alpha}\left[1+\mathcal{O}\left\{ \left|\Delta\tau^{*}\right|^{\Delta}\right\} \right]$
(ignoring critical constant and regular background terms), the asymptotic
behavior of the singular heat capacity is\begin{equation}
\Delta c_{V}^{*}=\frac{A^{\pm}}{\alpha}\left|\Delta\tau^{*}\right|^{-\alpha}\left[1+\alpha A_{1}^{\pm}\left|\Delta\tau^{*}\right|^{\Delta}\right]\label{Wegner eq cv (58)}\end{equation}
 Therefore, the leading amplitudes $A^{\pm}$ can be estimated from\begin{equation}
A^{\pm}=\left(Y_{c}\right)^{2-\alpha}\mathcal{Z}_{A}^{\pm}\label{A amplitude (59)}\end{equation}
which relates $A^{\pm}$ only to the single scale factor $Y_{c}$.
However the implicit role {[}see Eq. (\ref{Aparticle amplitude (56)})]
of the particle number $\frac{1}{Z_{c}}$ filling the critical interaction
cell, cannot be ignored for basic understanding of the master thermophysical
properties of the one-component fluid subclass.

As expected, using Eqs. (\ref{sigma0 amplitude (45)}) and (\ref{A amplitude (59)}),
we retrieve the universal amplitude combinations of Eqs. (\ref{universal RsigmaA (50)}),
such as\begin{equation}
R_{\sigma A}^{\pm}=\beta_{c}\sigma_{0}\left(A^{\pm}\right)^{\frac{d-1}{d}}=\mathcal{Z}_{\sigma}\left(\mathcal{Z}_{A}^{\pm}\right)^{\frac{d-1}{d}}\label{universal RsigmaA (60)}\end{equation}
Simultaneously, using Eqs. (\ref{ksi0 amplitude (47)}) and (\ref{A amplitude (59)}),
we also retrieve the well-known universal quantities\begin{equation}
\begin{array}{cl}
\mathcal{A}_{S}^{*}\times\left(\ell^{*}\right)^{d} & \sim\left(\mathcal{T}^{*}\right)^{2}\times\mathcal{C}_{S}^{*}\times\left(\ell^{*}\right)^{d}=\left(R_{\xi}^{\pm}\right)^{d}\\
 & =\left[\mathcal{Z}_{\sigma}\left(\mathcal{Z}_{A}^{\pm}\right)^{\frac{1}{d}}\right]^{d}\\
 & =\left[\xi^{\pm}\left(A^{\pm}\right)^{\frac{1}{d}}\right]^{d}\end{array}\label{universal Rksi (61)}\end{equation}
where $R_{\xi}^{+}\approxeq0.2696$ and $R_{\xi}^{-}\approxeq0.1692$,
for $d=3$ \citet{Bagnuls2002}.

Summarizing the above results for (seven) singular behaviors {[}surface
tension, ($\pm$)-correlation length, ($\pm$)-heat capacity, and
($\pm$)-isothermal susceptibility], we note that the (five) Eqs.
(\ref{B amplitude (44)}), (\ref{sigma0 amplitude (45)}), (\ref{S0 amplitude (46)}),
(\ref{ksi0 amplitude (47)}), and (\ref{Wegner eq cv (58)}) \emph{close}
the hyperscaling universal features \emph{along the critical isochore
above and below} $T_{c}$, in conformity with the two-scale-factor
universality. Therefore, among the three universal exponents $\nu$,
$\phi$, and $\alpha$, only one is readily independent {[}see Eqs.
(\ref{Widom scaling law (48)}) and (\ref{hyperscallaw dnualpha (52)})].
The related master - physical amplitudes $\mathcal{Z}_{\sigma}$ -
$\sigma_{0}$, $\mathcal{Z}_{\xi}^{\pm}$ - $\xi^{\pm}$, and $\mathcal{Z}_{A}^{\pm}$
- $A^{\pm}$, depend on uniquely $Y_{c}$ {[}see Eqs. (\ref{sigma0 amplitude (45)}),
(\ref{ksi0 amplitude (47)}) and (\ref{A amplitude (59)})].

As a partial but essential conclusion, amplitude $\sigma_{0}$ of
the interfacial tension, is only characterized by the scale factor
$Y_{c}$ accounting for the nonuniversal microscopic nature of each
fluid crossing its critical point along the critical isochore.

\subsection{The order parameter density dependence characterized by the $Z_c$ scale factor}

The use of Eq. (\ref{master sigma vs mop (35)}) to estimate the effective
parameters of Eqs. (\ref{parachor correlation (1)}) or (\ref{sigma vs op (10)}),
leads to the Ising-like expressions for the parachor exponent,\begin{equation}
\pi_{a}=\frac{\phi}{\beta},\label{parachor exponent (62)}\end{equation}
and the (asymptotical) parachor,\begin{equation}
\begin{array}{cl}
P_{0} & =\frac{M_{mol}}{2\rho_{c}}\frac{\left(Z_{c}\right)^{\frac{1}{2}}\left[\alpha_{c}p_{c}\mathcal{Z}_{\sigma}\right]^{\pi_{a}}}{\mathcal{Z}_{M}}\\
 & =\left(Z_{c}\right)^{\frac{3}{2}}\left(\alpha_{c}\right)^{d-\left(d-1\right)\frac{\phi}{\beta}}\left\{ \frac{\left[\left(\beta_{c}\right)^{-1}\mathcal{Z}_{\sigma}\right]^{\pi_{a}}}{2\mathcal{Z}_{M}}\right\} ,\end{array}\label{parachor amplitude (63)}\end{equation}
In spite of the complex combination of scale factors, we note that
$Y_{c}$ does not appear in the right hand side of Eq. (\ref{parachor amplitude (63)}).
However, as clearly shown in the above § 3.1, $Y_{c}$ is the characteristic
scale factor of the critical isochoric path where interfacial properties
are defined. This amazing and important result is entirely due to
the \emph{hyperscaling} universal feature \emph{associated to the
critical isothermal path}, as will be discussed below.

Indeed, to close the discussion on hyperscaling in critical phenomena,
we need to introduce the two universal exponents $\eta\approx0.035$
and $\delta\approx4.82$, characterizing universal features of correlation
function at the critical point and thermodynamic function along the
critical isotherm, respectively \citet{hyperscaling}. $\eta$ and
$\delta$ are related by the hyperscaling law\begin{equation}
\frac{2-\eta}{d}=\frac{\delta-1}{\delta+1}\label{hyperscallaw detadelta (64)}\end{equation}
Equation (\ref{hyperscallaw detadelta (64)}), added to the previous
Eq. (\ref{hyperscallaw dnualpha (52)}), relate in a unequivocal manner
the two (independent) exponents $\delta$ and $\alpha$ describing
the thermodynamics, and the two (independent) exponents $\eta$ and
$\nu$ describing the correlations, via $d$ uniquely. In addition,
each \{thermodynamic-correlation\} exponent pair, either $\left\{ \delta;\eta\right\} $,
or $\left\{ \alpha;\nu\right\} $, characterizes each independent
thermodynamic path to reach the critical point, either the critical
isothermal line and the critical point itself, or the critical isochoric
line, respectively \citet{Garrabos1985}.

Obviously, universal values of the corresponding amplitude combinations
have been theoretically estimated \citet{Privman1991}. We have already
given the universal amplitude combination$\left(R_{\xi}^{\pm}\right)^{d}=A^{\pm}\left(\xi^{\pm}\right)^{d}$
(valid along the critical isochore), associated to the scaling law
$d\nu=2-\alpha$. So that, to close the presentation of the two-scale-factor
universality, we can also consider the universal amplitude combination\begin{equation}
R_{D}=D_{\rho}^{c}\left(\widehat{D}_{\rho}\right)^{\frac{\delta+1}{2}}=D_{n}^{c}\left(\widehat{D}_{n}\right)^{\frac{\delta+1}{2}}\label{universal RD (65)}\end{equation}
along the critical isotherm, associated to the hyperscaling law of
Eq. (\ref{hyperscallaw detadelta (64)}). Here, we have anticipated
(see below) the introduction of the leading amplitudes $D_{\rho}^{c}$
($D_{n}^{c}$) and $\widehat{D}_{\rho}$ ($\widehat{D}_{n}$) associated
to the singular shape of the ordering field along the critical isotherm
and to the singular decreasing of the correlation function at the
critical point, respectively. In the $D_{x}^{y}$, $\widehat{D}_{x}$
notations, i) superscript $y=c$ recalls for the non-zero value of
the order parameter density in a fluid maintained at constant critical
temperature; ii) decorated hat recalls for the infinite size of the
order parameter density fluctuations in a critical fluid maintained
exactly at the critical point; iii) subscript $x=\rho$ recalls for
a thermodynamic potential which is normalized per volume unit and
a definition of the order parameter density related to the mass density,
namely $\Delta\tilde{\rho}=\frac{\rho-\rho_{c}}{\rho_{c}}$ (then
$D_{\rho}^{c}\equiv D$, where $D$ is the customary notation of this
leading amplitude); and iv) subscript $x=n$ recalls for a thermodynamic
potential which is normalized per particle and a definition of the
order parameter density related to the number density, namely $\Delta m^{*}=\left(\alpha_{c}\right)^{d}\left(n-n_{c}\right)$.

Considering the \{thermodynamic-correlation\} pairs $\left\{ D_{\rho}^{c};\widehat{D}_{\rho}\right\} $,
or $\left\{ D_{n}^{c};\widehat{D}_{n}\right\} $ defined along the
critical isotherm and at the critical point, and the \{thermodynamic-correlation\}
pair $\left\{ A^{\pm};\xi^{\pm}\right\} $ defined along the critical
isochore, from Eqs. (\ref{universal Rksi (61)}) and (\ref{universal RD (65)}),
a single amplitude characterizes each thermodynamic path crossing
the critical point, either at constant critical temperature, or at
constant critical density, respectively. Considering the \{interfacial-bulk\}
pairs $\left\{ \sigma_{0};A^{\pm}\right\} $ and $\left\{ \sigma_{0};\xi^{\pm}\right\} $,
the previous section has shown that $Y_{c}$ is precisely the single
scale factor of the temperature field which characterizes the critical
isochore. Therefore, in conformity with the two-scale-factor universality,
we are now concerned by the existence of the equivalent \{interfacial-bulk\}
pairs, which should involve Ising-like leading mplitude of the parachor
correlations and either $D_{\rho}^{c}$ or $\widehat{D}_{\rho}$.
Closing their respective $Z_{c}$-dependence demonstrates then that
$Z_{c}$ is precisely the single scale factor of the ordering field
which characterizes the critical isotherm. Obviously, the above Eq.
(\ref{parachor amplitude (63)}) where $P_{0}$ appears only $Z_{c}$-dependent
is already in agreement with such an universal feature.

Starting with the scaling law,\begin{equation}
\beta\left(\delta+1\right)=2-\alpha\label{betadeltaalpha scallaw (66)}\end{equation}
and using the hyperscaling law of Eq. (\ref{hyperscallaw detadelta (64)}),
we can recalculate $\pi_{a}=\frac{\phi}{\beta}$ {[}see Eq. (\ref{parachor exponent (62)}].
We obtain\begin{equation}
\frac{\pi_{a}}{d-1}=\frac{2}{d-\left(2-\eta\right)}=\frac{\delta+1}{d}\label{parachor hyperscallaw(67)}\end{equation}
with $d=3$. The unequivocal link between the Ising-like parachor
exponent $\pi_{a}$ and either $\eta$, or $\delta$, is now undoubtedly
due to \emph{mixed hyperscaling} \emph{along the critical isotherm
and at the critical point itself}, expliciting the respective interface
($d-1$) and bulk ($d$) dimensions. This Eq. (\ref{parachor hyperscallaw(67)})
completes the similar \emph{mixed hyperscaling} link $\frac{\phi}{d-1}=\nu=\frac{2-\alpha}{d}$
{[}see Eq. (\ref{Widom scaling law (48)})] between interfacial exponent
$\phi$ and either $\nu$, or $\alpha$, \emph{along the critical
isochore}.

To find the universal amplitude combinations associated with Eq. (\ref{parachor hyperscallaw(67)}),
we need to introduce an unambiguous definition of the parachor correlations
from the corresponding Wegner-like expansions expressed in terms of
the order parameter density. The following (master and physical) power
laws\begin{equation}
\Sigma^{*}=\mathcal{\widetilde{Z}}_{\sigma}\left(\mathcal{M}_{LV}^{*}\right)^{\pi_{a}}\left\{ 1+\mathcal{O}\left[\left(\mathcal{M}_{LV}^{*}\right)^{\frac{\Delta}{\beta}}\right]\right\} \label{gsigma master asymplaw (68)}\end{equation}
\begin{equation}
\begin{array}{cl}
\sigma^{*} & =D_{n}^{\sigma}\left(\Delta m_{LV}^{*}\right)^{\pi_{a}}\left\{ 1+\mathcal{O}\left[\left(\Delta m_{LV}^{*}\right)^{\frac{\Delta}{\beta}}\right]\right\} \\
 & =D_{\rho}^{\sigma}\left(\Delta\tilde{\rho}_{LV}\right)^{\pi_{a}}\left\{ 1+\mathcal{O}\left[\left(\Delta\tilde{\rho}_{LV}\right)^{\frac{\Delta}{\beta}}\right]\right\} \end{array}\label{sigmastar asymplaw (69)}\end{equation}
are more appropriate than Eqs. (\ref{parachor correlation (1)}),
or (\ref{sigma vs op (10)}) in the sense where Eq. (\ref{gsigma master asymplaw (68)})
{[}or (\ref{sigmastar asymplaw (69)})] acts as a two-dimensional
equation of state for the liquid-vapor interface (along the critical
isochore). The dimensionless amplitudes $D_{n}^{\sigma}$ and $D_{\rho}^{\sigma}$
are called Ising-like parachors to distinguish them from dimensional
$P_{0}$ {[}see Eq. (\ref{parachor amplitude (63)})] called parachor.
We recall that Eq. (\ref{sigmastar asymplaw (69)}) refers to the
dimensionless interfacial tension, $\sigma^{*}\equiv\frac{\sigma}{\alpha_{c}p_{c}}$.
Now, the superscript $y=\sigma$ in $D_{x}^{y}$ notations, recalls
for the thermodynamic definition of the interfacial tension of a non-homogeneous
fluid where the order parameter density spontaneously is non-zero,
along the critical isochore. As mentioned above, $x=n$ and $x=\rho$
reflect the two forms $\Delta m^{*}$ and $\Delta\tilde{\rho}$ of
the order parameter density, leading to the r.h.s. forms of Eq. (\ref{sigmastar asymplaw (69)}).
The related $Z_{c}$-dependence between $D_{n}^{\sigma}$ and $D_{\rho}^{\sigma}$
is\begin{equation}
D_{n}^{\sigma}=\left(Z_{c}\right)^{\pi_{a}}D_{\rho}^{\sigma}\label{Dsigma-Drhosigma vs Zc (70)}\end{equation}

Using Eqs. (\ref{master ordering field (27)}) and (\ref{master surface tension (29)})
to compare the leading terms of master and physical Eqs. (\ref{gsigma master asymplaw (68)})
and (\ref{sigmastar asymplaw (69)}), we obtain

\begin{equation}
\begin{array}{lll}
D_{n}^{\sigma}=\left(Z_{c}\right)^{\frac{d}{2}\pi_{a}}\mathcal{\widetilde{Z}}_{\sigma} & =\left(Z_{c}\right)^{\delta+1}\mathcal{\widetilde{Z}}_{\sigma} & =\left(Z_{c}\right)^{\frac{d\left(d-1\right)}{d-2+\eta}}\mathcal{\widetilde{Z}}_{\sigma}\\
D_{\rho}^{\sigma}=\left(Z_{c}\right)^{\frac{\pi_{a}}{2}}\mathcal{\widetilde{Z}}_{\sigma} & =\left(Z_{c}\right)^{\frac{\delta+1}{d}}\mathcal{\widetilde{Z}}_{\sigma} & =\left(Z_{c}\right)^{\frac{d-1}{d-2+\eta}}\mathcal{\widetilde{Z}}_{\sigma}\end{array}\label{Dsigma vs Zc (71)}\end{equation}
 As expected, Eqs. (\ref{Dsigma vs Zc (71)}) provide unequivocal
determinations of $D_{n}^{\sigma}$ and $D_{\rho}^{\sigma}$ from
the scale factor $Z_{c}$ (selecting either $\pi_{a}$, or $\delta$,
or $\eta$, as independent exponent).

We can define in a similar manner the $Z_{c}$-dependence of $D_{x}^{c}$
and $\widehat{D}_{x}$ introduced through Eq. (\ref{universal RD (65)}).

First, the amplitude $D_{x}^{c}$ are associated to the singular behavior
of the ordering field in a three-dimensional homogeneous fluid in
contact with a particle reservoir, fixing the non-zero value of the
order parameter density, and thermostated at constant (critical) temperature
$T=T_{c}$. In that thermodynamic situation, it is established that
the (master and physical) ordering fields obey the following power
laws\begin{equation}
\mathcal{H}^{*}=\pm\mathcal{Z}_{H}^{c}\left|\mathcal{M}^{*}\right|^{\delta}\left\{ 1+\mathcal{O}\left[\left|\mathcal{M}^{*}\right|^{\frac{\Delta}{\beta}}\right]\right\} \label{ordering field meos (72)}\end{equation}
\begin{equation}
\begin{array}{cc}
\Delta\mu_{\bar{p}}^{*} & =\pm D_{n}^{c}\left|\Delta m^{*}\right|^{\delta}\left\{ 1+\mathcal{O}\left[\left|\Delta m^{*}\right|^{\frac{\Delta}{\beta}}\right]\right\} \\
\Delta\tilde{\mu}_{\rho} & =\pm D_{\rho}^{c}\left|\Delta\tilde{\rho}\right|^{\delta}\left\{ 1+\mathcal{O}\left[\left|\Delta\tilde{\rho}\right|^{\frac{\Delta}{\beta}}\right]\right\} \end{array}\label{ordering field eos (73)}\end{equation}
where $\mathcal{Z}_{H}^{c}\approx252$ is a master value for the one-component
fluid subclass. As in the above case of dimensionless interfacial
tension, the r.h.s. forms of Eqs. (\ref{ordering field eos (73)})
refer to distinct order parameter densities, $\Delta m^{*}$ and $\Delta\tilde{\rho}$,
leading to the following $Z_{c}$-dependence between $D_{n}^{c}$
and $D_{\rho}^{c}$\begin{equation}
D_{n}^{c}=\left(Z_{c}\right)^{\delta}D_{\rho}^{c}=\left(Z_{c}\right)^{\frac{d+2-\eta}{d-2+\eta}}D_{\rho}^{c}\label{Dc-Drhoc vs Zc (74)}\end{equation}
 In the case of a critical isothermal fluid, a convenient rewriting
of the leading term in  Eq. (\ref{ordering field eos (73)}) is \citet{Levelt1981}
\begin{equation}
\left(\mu_{\rho}-\mu_{\rho,c}\right)\frac{\rho_{c}}{p_{c}}=\frac{p-p_{c}}{p_{c}}=\pm D_{\rho}^{c}\left|\Delta\rho^{*}\right|^{\delta}\label{physical isotherm (75)}\end{equation}
 Using Eqs. (\ref{master ordering field (27)}) and (\ref{master order parameter (28)})
and accounting for dual definitions of the ordering field - order
parameter density with respect to appropriate free energies, the comparison
of the leading terms in Eqs. (\ref{ordering field meos (72)}) and
(\ref{ordering field eos (73)}) gives the following results\begin{equation}
\begin{array}{cc}
D_{n}^{c}=\left(Z_{c}\right)^{\frac{d}{2}\left(\delta+1\right)}\mathcal{Z}_{H}^{c} & =\left(Z_{c}\right)^{\frac{d^{2}}{d-2+\eta}}\mathcal{Z}_{H}^{c}\\
D_{\rho}^{c}=\left(Z_{c}\right)^{\frac{\delta+1}{2}}\mathcal{Z}_{H}^{c} & =\left(Z_{c}\right)^{\frac{d}{d-2+\eta}}\mathcal{Z}_{H}^{c}\end{array}\label{Disotherm vs Zc (76)}\end{equation}
 Selecting then either $\delta$, or $\eta$, as an independent exponent,
Eqs. (\ref{Disotherm vs Zc (76)}) relate unequivocally each respective
physical amplitude $D_{n}^{c}$ or $D_{\rho}^{c}$, to the scale factor
$Z_{c}$.

Second, the amplitude $\widehat{D}_{x}$ is associated to the singular
behavior of the dimensionless spatial correlation function $G^{*}\left(\Delta\tau^{*}=0,\Delta x^{*}=0,r^{*}=\frac{r}{\alpha_{c}}\right)\propto\left(\frac{1}{r^{*}}\right)^{\frac{1}{d-2+\eta}}$
at the exact critical point ($r$ is the direct space position, $x=n$
or $x=\rho$ following the order parameter density choice). More precisely,
introducing the static structure factor $\chi_{x}\left(T-T_{c},x-x_{c},q\right)$,
where $q$ is the wavenumber in the reciprocal space, such as $\chi_{x}\left(T-T_{c},x-x_{c},0\right)$
takes the same dimension as the corresponding isothermal susceptibility
(see below and reference ), we define the following dimensionless
singular form of the master and physical structure factors \begin{equation}
\mathcal{X}^{*}\left(\mathcal{T}^{*}=0,\mathcal{M}^{*}=0,\mathcal{Q}^{*}\equiv q^{*}\right)\propto\widehat{\mathcal{Z}}_{G}\left(\mathcal{Q}^{*}\right)^{\eta-2}\label{master structure factor (77)}\end{equation}
\begin{equation}
\begin{array}{rl}
\chi_{n}^{*}\left(\Delta\tau^{*}=0,\Delta m^{*}=0,q^{*}=q\alpha_{c}\right) & \propto\widehat{D}_{n}\left(q^{*}\right)^{\eta-2}\\
\chi_{\rho}^{*}\left(\Delta\tau^{*}=0,\Delta\tilde{\rho}=0,q^{*}=q\alpha_{c}\right) & \propto\widehat{D}_{\rho}\left(q^{*}\right)^{\eta-2}\end{array}\label{physical structure facor (78)}\end{equation}
where $\widehat{\mathcal{Z}}_{G}$ is a master constant for the one-component
fluid subclass. Starting from the isothermal susceptibilities $\chi_{T,n}=\left(\frac{\partial n}{\partial\mu_{\bar{p}}}\right)_{T}$
and $\chi_{T,\rho}=\left(\frac{\partial\rho}{\partial\mu_{\rho}}\right)_{T}$,
associated to the order parameter densities $\Delta m^{*}$ and $\Delta\tilde{\rho}$,
respectively, it is easy to obtain the following relation between
the amplitudes of the right hand side of Eq. (\ref{physical structure facor (78)})
\begin{equation}
\widehat{D}_{n}=\left(Z_{c}\right)^{-2}\widehat{D}_{\rho}\label{Dnrhochapeau vs Zc (79)}\end{equation}
Similarly, using Eqs. (\ref{master structure factor (77)}), (\ref{physical structure facor (78)}),
and adding the relations between the master isothermal susceptibility
$\mathcal{X}^{*}=\left(\frac{\partial\mathcal{M}^{*}}{\partial\mathcal{H}^{*}}\right)_{\mathcal{T}^{*}}$
and their associated physical dimensionless forms (neglecting quantum
effects), we obtain the following relations \begin{equation}
\begin{array}{c}
\left(Z_{c}\right)^{d}\widehat{D}_{n}=\widehat{\mathcal{Z}}_{G}\\
Z_{c}\widehat{D}_{\rho}=\widehat{\mathcal{Z}}_{G}\end{array}\label{Dhat vs Zc (80)}\end{equation}
Each one of Eqs. (\ref{Dhat vs Zc (80)}) gives the expected unequivocal
link between $\widehat{D}_{n}$, or $\widehat{D}_{\rho}$, and $Z_{c}$.
We note that $\widehat{D}_{n}$, or $\widehat{D}_{\rho}$, and $Z_{c}$
are true critical numbers, i.e. dimensionless quantities defined at
the critical point, exactly. One among these critical numbers characterizes
the selected one-component fluid. Therefore, the above link is {}``basic''
because it only depends of the hypothesized linear relation between
master and physical conjugated (ordering field-order parameter density)
variables. Eliminating $Z_{c}$ between Eqs. (\ref{Disotherm vs Zc (76)})
and (\ref{Dhat vs Zc (80)}) provides the universal amplitude combination
of Eq. (\ref{universal RD (65)}), which closes the universal features
along the critical isotherm and at the exact critical point, in conformity
with the two-scale-factor universality. Finally, Eqs. (\ref{Aparticle amplitude (56)})
and (\ref{Dhat vs Zc (80)}) are the necessary \emph{closure equations}
which unequivocally relates the two (independent) leading amplitudes
and the two (independent) scale factors characteristics of each one-component
fluid, selecting $\eta$ and $\alpha$ as two (independent) critical
exponents.

>From Eq. (\ref{Dsigma vs Zc (71)}) and Eqs. (\ref{Disotherm vs Zc (76)})
or (\ref{Dhat vs Zc (80)}), associated with hyperscaling law of Eq.
(\ref{parachor hyperscallaw(67)}), it is immediate to construct the
following new combinations between \emph{interfacial} amplitudes and
\emph{bulk} amplitudes, whose values are expected to be universal\begin{equation}
R_{D\sigma}=\frac{D_{n}^{c}}{\left(D_{n}^{\sigma}\right)^{\frac{d}{d-1}}}=\frac{D_{\rho}^{c}}{\left(D_{\rho}^{\sigma}\right)^{\frac{d}{d-1}}}=\frac{\mathcal{Z}_{H}^{c}}{\left(\mathcal{\widetilde{Z}}_{\sigma}\right)^{\frac{d}{d-1}}}\label{RDsigma (81)}\end{equation}
\begin{equation}
\begin{array}{cl}
R_{\widehat{D}\sigma} & =\widehat{D}_{n}\left(D_{n}^{\sigma}\right)^{\frac{d-2+\eta}{d-1}}=\widehat{D}_{\rho}\left(D_{\rho}^{\sigma}\right)^{\frac{d-2+\eta}{d-1}}\\
 & =\widehat{\mathcal{Z}}_{G}\left(\mathcal{\widetilde{Z}}_{\sigma}\right)^{\frac{d-2+\eta}{d-1}}\end{array}\label{RDchapeausigma (82)}\end{equation}

To complete our understanding of the universal features related to
a (constrained or spontaneous) non-zero value of the order parameter
density, we must compare also the \emph{bulk} properties of each (liquid-like
or gas-like) single phase at critical temperature $T=T_{c}$, and
the \emph{bulk} properties of each (liquid or gas) coexisting phase
in the non-homogenous domain $T<T_{c}$ (admitting then a \emph{symmetrical}
one-component fluid close to the critical point). In these comparable
three-dimensional situations where the symmetrized order parameter
density can take the same non-zero value at two different temperatures,
the existence of universal proportionality (in units of $\left(\beta_{c}\right)^{-1}$)
is expected for the singular bulk free energy $E^{*}$ of a homogeneous
phase, either maintained at constant (critical) temperature $T=T_{c}$
{[}i.e., $\frac{\text{bulk}\,\text{free}\,\text{energy}}{V}\propto D_{\rho}^{c}\left|\Delta\tilde{\rho}\right|^{\delta+1}\propto D_{n}^{c}\left|\Delta m^{*}\right|^{\delta+1}$],
or at constant (critical) volume at $T$ below $T_{c}$ {[}i.e., $\frac{\text{bulk}\,\text{free}\,\text{energy}}{V_{L,V}}\propto\frac{A^{-}}{\alpha\left(1-\alpha\right)\left(2-\alpha\right)}\left|\Delta\tau^{*}\right|^{2-\alpha}$].
We account then for the thermodynamic constraints for coexisting phases,
expressed by $\left|\Delta\tau^{*}\right|\propto\left(\frac{\left|\Delta\tilde{\rho}\right|}{B_{\rho}}\right)^{\frac{1}{\beta}}\propto\left(\frac{\left|\Delta m^{*}\right|}{B_{n}}\right)^{\frac{1}{\beta}}$,
and for the universal features above and below the critical temperature
along the critical isochore, expressed by the universal ratio $\frac{A^{+}}{A^{-}}\simeq0.537$.
As a result, we obtain the following universal amplitude combinations
{[}with $B_{\rho}\equiv B$ and $B_{n}=\left(Z_{c}\right)^{-1}B$,
where $B$ is the customary notation of this leading amplitude, see
Eq. (\ref{Wegner eq deltarho (4)})] \begin{equation}
\left(Q_{B}^{\pm}\right)^{\delta+1}=\frac{B^{\delta+1}}{A^{\pm}}D_{\rho}^{c}=\frac{\left(B_{n}\right)^{\delta+1}}{A^{\pm}}D_{n}^{c}\label{QB (83)}\end{equation}
This amplitude combination is related to the ''cross'' scaling laws
{[}see Eqs. (\ref{cross scalaw dmualpha (51)}) and (\ref{betadeltaalpha scallaw (66)})]
\begin{equation}
\frac{d\nu}{\beta}=\frac{2-\alpha}{\beta}=\delta+1\label{alphabetadelta scallaw (84)}\end{equation}
Hereabove, Eq. (\ref{alphabetadelta scallaw (84)}) combines exponent
ratios $\frac{\nu}{\beta}$ or $\frac{2-\alpha}{\beta}$, which caracterize
\emph{bulk} properties expressed as a function of the order parameter
density in the nonhomogeneous domain, to the exponent $\delta$ which
caracterizes the ordering field as a function of the order parameter
density along the critical isotherm. Equations (\ref{RDsigma (81)})
and (\ref{QB (83)}) imply that the Ising-like parachors $D_{\rho}^{\sigma}$
(or $D_{n}^{\sigma}$, equivalently), can also be expressed in terms
of the ratio $\frac{B^{\delta+1}}{A^{\pm}}$ (or $\frac{\left(B_{n}\right)^{\delta+1}}{A^{\pm}}$,
equivalently), eliminating then $D_{\rho}^{c}$ (or $D_{\rho}^{c}$,
equivalently). As a matter of fact, despite an explicit $Y_{c}$ dependence
in the amplitudes $B\propto\left(Z_{c}\right)^{\frac{1}{2}}\left(Y_{c}\right)^{\beta}$
and $A^{\pm}\propto\left(Y_{c}\right)^{2-\alpha}\mathcal{Z}_{A}^{\pm}$,
their ratio $\frac{B^{\delta+1}}{A^{\pm}}$ always takes appropriate
forms to ensure the disappearance of the $Y_{c}$-scale factor, and
only reflect hyperscaling attached to the critical isotherm, which
is characterized by the $Z_{c}$-scale factor, uniquely. As a consequence,
we obtain the universal combinations \begin{equation}
\begin{array}{cl}
\frac{\left(Q_{B}^{\pm}\right)^{\delta+1}}{R_{D\sigma}} & =\left(D_{n}^{\sigma}\right)^{\frac{d}{d-1}}\frac{\left(B_{n}\right)^{\delta+1}}{A^{\pm}}=\left(D_{\rho}^{\sigma}\right)^{\frac{d}{d-1}}\frac{B^{\delta+1}}{A^{\pm}}\\
 & =\left(\mathcal{\widetilde{Z}}_{\sigma}\right)^{\frac{d}{d-1}}\frac{\left(\mathcal{Z}_{M}\right)^{\delta+1}}{\mathcal{Z}_{A}^{\pm}}\end{array}\label{QBvsRDsigma (85)}\end{equation}
Similarly, we note that the amplitude products $\Gamma^{\pm}B^{\delta-1}$
or $\Gamma_{n}^{\pm}\left(B_{n}\right)^{\delta-1}$ are associated
to the {}``cross'' scaling laws \begin{equation}
\frac{d\nu}{\gamma\left(\delta+1\right)}=\frac{\beta}{\gamma}=\frac{1}{\delta-1}\label{betagammadelta scallaw (86)}\end{equation}
which also reflect hyperscaling attached to the critical isotherm.
Here above, $\Gamma^{\pm}\equiv\Gamma_{\rho}^{\pm}$ and $\Gamma_{n}^{\pm}=\left(Z_{c}\right)^{-2}\Gamma^{\pm}$
are the leading amplitudes of the singular behavior of $\tilde{\chi}_{T,\rho}$
and $\chi_{T,n}^{*}$, while $\gamma\simeq1.24$ is the related critical
exponent {[}where $\Gamma^{\pm}$ are the customary notations along
the critical isochore, see below Eq. (\ref{khiTrho-khiTn (87)}].
The (physical) dimensionless susceptibilities obey the following power
laws\begin{equation}
\begin{array}{cc}
\tilde{\chi}_{T,\rho} & =\Gamma^{\pm}\left|\Delta\tau^{*}\right|^{-\gamma}\left[1+\overset{i=\infty}{\underset{i=1}{\sum}}\Gamma_{i}^{\pm}\left|\Delta\tau^{*}\right|^{i\Delta}\right]\\
\chi_{T,n}^{*} & =\Gamma_{n}^{\pm}\left|\Delta\tau^{*}\right|^{-\gamma}\left[1+\overset{i=\infty}{\underset{i=1}{\sum}}\Gamma_{i}^{\pm}\left|\Delta\tau^{*}\right|^{i\Delta}\right]\end{array}\label{khiTrho-khiTn (87)}\end{equation}
The corresponding (two-term) singular behavior of the master susceptibility
$\mathcal{X}^{*}=\left(Z_{c}\right)^{d}\chi_{T,n}^{*}=Z_{c}\tilde{\chi}_{T,\rho}$,
is\begin{equation}
\mathcal{X}^{*}=\mathcal{Z}_{\chi}^{\pm}\left|\mathcal{T}^{*}\right|^{-\gamma}\left[1+\mathcal{Z}_{\chi}^{1,\pm}\left|\mathcal{T}^{*}\right|^{\Delta}+...\right]\label{Masterkhi (88)}\end{equation}
where $\mathcal{Z}_{\chi}^{+}\simeq0.119$ and $\mathcal{Z}_{\chi}^{-}\simeq0.0248$
are the master values of leading amplitudes, with universal ratio
$\frac{\mathcal{Z}_{\chi}^{+}}{\mathcal{Z}_{\chi}^{-}}=\frac{\Gamma^{+}}{\Gamma^{-}}\simeq4.79$
\citet{Guida1998}. Now, the explicit $Y_{c}$-dependence $B\propto\left(Z_{c}\right)^{\frac{1}{2}}\left(Y_{c}\right)^{\beta}$
and $\Gamma^{\pm}\propto\left(Z_{c}\right)^{-x}\left(Y_{c}\right)^{-\gamma}\mathcal{Z}_{\chi}^{\pm}$
desappear in their combination $\Gamma^{\pm}B^{\delta-1}$, due to
Eq. (\ref{betagammadelta scallaw (86)}). This latter product reflects
hyperscaling attached to the $Z_{c}$-scale factor of the critical
isotherm, uniquely. Introducing the universal combination\begin{equation}
R_{\chi}=\Gamma^{+}B^{\delta-1}D=\Gamma_{n}^{+}\left(B_{n}\right)^{\delta-1}D_{n}^{c}\label{Rkhi (89)}\end{equation}
 to eliminate $D_{\rho}^{c}\equiv D$ or $D_{n}^{c}$ using Eqs. (\ref{RDsigma (81)})
and (\ref{Rkhi (89)}), we obtain the following universal combinations
which contain the Ising-like parachors\begin{equation}
\begin{array}{cl}
\frac{R_{\chi}}{R_{D\sigma}} & =\left(D_{n}^{\sigma}\right)^{\frac{d}{d-1}}\Gamma_{n}^{+}\left(B_{n}\right)^{\delta-1}\\
 & =\left(D_{\rho}^{\sigma}\right)^{\frac{d}{d-1}}\Gamma^{+}B^{\delta-1}=\left(\mathcal{\widetilde{Z}}_{\sigma}\right)^{\frac{d}{d-1}}\mathcal{Z}_{\chi}^{+}\left(\mathcal{Z}_{M}\right)^{\delta-1}\end{array}\label{RkhivsRDsigma (90)}\end{equation}

Summarizing the results for the above (seven) singular properties
(surface tension, order parameter density, ordering field, ($\pm$)-heat
capacity, and ($\pm$)-isothermal susceptibility), we note that (three)
Eqs. (\ref{universal RD (65)}), (\ref{QB (83)}), (\ref{Rkhi (89)}),
and (two) universal ratios $\frac{\mathcal{Z}_{A}^{+}}{\mathcal{Z}_{A}^{-}}=\frac{A^{+}}{A^{-}}$,
$\frac{\mathcal{Z}_{\chi}^{+}}{\mathcal{Z}_{\chi}^{-}}=\frac{\Gamma^{+}}{\Gamma^{-}}=\frac{\Gamma_{n}^{+}}{\Gamma_{n}^{-}}$,
close the hyperscaling universal features, \emph{at the critical point,
along the critical isotherm, and in the nonhomogeneous domain}, in
conformity with the two-scale-factor universality. Therefore, among
the exponents $\eta$, $\pi_{a}$, $\delta$, and exponent ratios
$\frac{2-\alpha}{\beta}$, $\frac{\gamma}{\beta}$, only one is readily
independent {[}see Eqs. (\ref{hyperscallaw detadelta (64)}), (\ref{parachor hyperscallaw(67)}),
(\ref{alphabetadelta scallaw (84)}), and (\ref{betagammadelta scallaw (86)})].
The related master - physical amplitudes $\mathcal{\widetilde{Z}}_{\sigma}$
- $D_{x}^{\sigma}$, $\mathcal{Z}_{H}^{c}$ - $D_{x}^{c}$, $\widehat{\mathcal{Z}}_{G}$
-$\widehat{D}_{x}$, and the related master - physical combinations
$\frac{\mathcal{Z}_{A}^{\pm}}{\left(\mathcal{Z}_{M}\right)^{\delta+1}}$
- $\frac{A^{\pm}}{B^{\delta+1}}$, and $\mathcal{Z}_{\chi}^{\pm}\left(\mathcal{Z}_{M}\right)^{\delta-1}$
- $\Gamma^{\pm}B^{\delta-1}$, are uniquely $Z_{c}$-dependent (see
Eqs. (\ref{Dsigma vs Zc (71)}), (\ref{Disotherm vs Zc (76)}), (\ref{RDsigma (81)}),
(\ref{QBvsRDsigma (85)}), and (\ref{RkhivsRDsigma (90)})].

As a partial but essential conclusion, the Ising-like parachor of
the interfacial tension, expressed as a function of the order-parameter
density, is only characterized by the scale factor $Z_{c}$ proper
to account for nonuniversal microscopic nature of each fluid at its
critical point or crossing them along the critical isotherm.

\section{Conclusions}

In contrast with all previous studies on the parachor correlations,
the present estimation of the behavior of interfacial tension as a
function of the density difference of the coexisting vapor and liquid
phases in the critical region, is made without adjustable parameter
when $Q_{c}^{min}=\left\{ \left(\beta_{c}\right),\alpha_{c},Z_{c},Y_{c}\right\} $
is known for a selected one-component fluid. 

The \emph{interfacial-bulk} universal features of exponent pairs,
$\left\{ \phi;\alpha\right\} $ and $\left\{ \phi;\nu\right\} $,
or amplitude pairs, $\left\{ \sigma_{0};A^{\pm}\right\} $, and $\left\{ \sigma_{0};\xi^{\pm}\right\} $,
indicate that the singularities of the surface tension, the (thermodynamic)
heat capacity, and the (correlation) length, expressed as a function
of the temperature field along the critical isochore are well-characterized
by a single characteristic scale factor. Using the scale dilatation
method, we have shown that this (fluid-dependent) scale factor is
$Y_{c}$, precisely. Similarly, the \emph{interfacial-bulk} universal
features of exponent pairs, $\left\{ \pi_{a};\delta\right\} $ and
$\left\{ \pi_{a};\eta\right\} $, or amplitude pairs, $\left\{ D_{x}^{\sigma};D_{x}^{c}\right\} $,
and $\left\{ D_{x}^{\sigma};\hat{D}_{x}\right\} $, indicate that
the singularities of the surface tension, the (thermodynamic) ordering
field and susceptibility, and the (correlation) length, expressed
as a function of the order parameter density along the critical isotherm
and in the nonhomogeneous domain, are well-characterized by a single
characteristic scale factor. Using the scale dilatation method, we
have also shown that this (fluid-dependent) scale factor is precisely
$Z_{c}$. Moreover, the desappearence of the isochoric scale factor
$Y_{c}$ in the estimation of the (Ising-like and effective) parachors,
is here well-understood in terms of hyperscaling. $Y_{c}$ and $Z_{c}$
are two independent characteristic numbers. They are fundamental points
for future developments of parachor correlations. Such results must
also be accounted for, in equations of the saturated vapor pressure
curve, the enthalpy of formation of the vapor-liquid interface, and,
more generally, in ancillary equations where adjustable parameters
can be estimated using a limited number of well-defined fluid-dependent
quantities including $Y_{c}$ and $Z_{c}$.

Since the present approach accounts for complete universal features
of critical phenomena, thanks to the scale dilatation method, in the
absence of theoretical prediction for the surface tension, Fig. \ref{fig04-parachor}
may also be useful for correlating interfacial properties with master
equation of the correlation length, incorporating a phenomenological
contribution of the confluent corrections to the asymptotic limit
here analyzed. As a special mention, the complete classical-to-critical
crossover predicted from the Field Theory framework can be used with
an exact knowledge of the density domain where the correlation length
and the interface thickness reach the order of magnitude of the microscopic
molecular interaction. When the two-phase fluid properties change
from the critical point to the triple point in such a controlled situation,
the introduction of supplementary parameters, either having crossover
nature (such as the crossover temperature for example), or having
empirical origin (such as the acentric factor, for example), should
then be made to descriminate the non-universal character proper to
each fluid system revealed from Fig. \ref{fig04-parachor}, at large
values of $\mathcal{M}_{LV}^{*}$ {[}or from Fig. 3 of Ref. \citet{Garrabos2002},
at large values of $\mathcal{T}^{*}$]. However, in all cases, any
supplementary parameter would be used in conformity to the above master
singular behavior of the one-component fluid subclass, for which the
two scale factors are now specified in terms of thermodynamic continuity
across the critical point \citet{Garrabos1982,Garrabos2005}. 

\appendix

\section{Parachor correlation in the non-homogeneous domain}

\begin{figure}
\includegraphics[width=80mm,height=160mm,keepaspectratio]{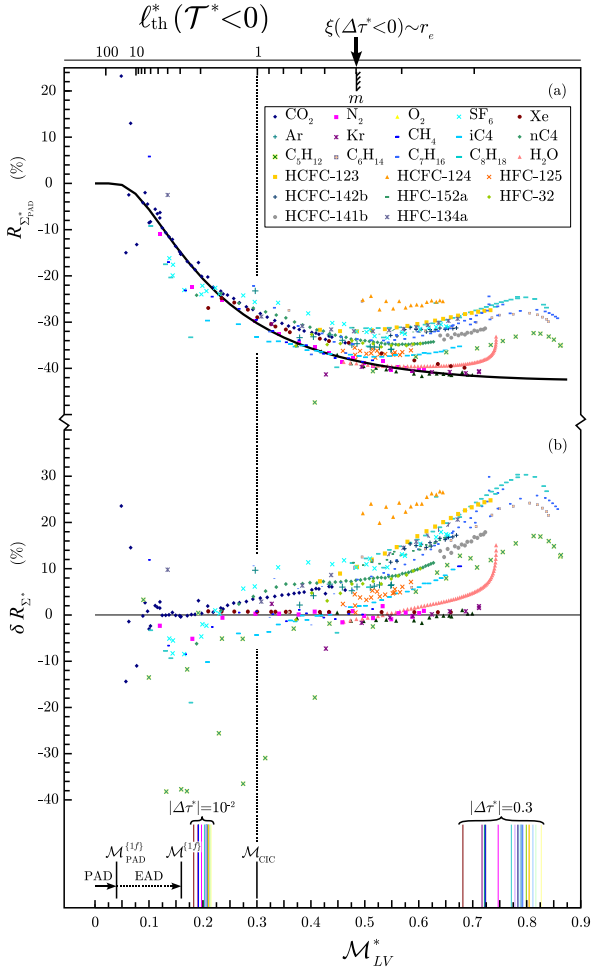}

\caption{(Color online) lin-lin scale: a) Residuals (expressed in \%) for the
experimental master parachor from reference to the asymptotical master
parachor calculated by Eq. (\ref{MastersigmaOP (36)}), as a function
of the master order parameter density; the full black curve corresponds
to Eq. (\ref{A2}) (see text); b) Deviation (expressed in \%) of the
experimental residuals estimated in part (a), from reference to the
full black curve of Eq. (\ref{A2}). \label{fig05}}

\end{figure}

We have shown that the power law $\mathcal{\widetilde{Z}}_{\sigma}\left(\mathcal{M}_{\text{th}}^{*}\right)^{\pi_{a}}$
{[}see Eq. (\ref{MastersigmaOP (36)})], where $\mathcal{M}_{\text{th}}^{*}$
can be estimated as a modified theoretical function of $\mathcal{T}^{*}<0$,
provides an asymptotic scaling behavior which agrees with the available
experimental results in the range $\mathcal{M}_{LV}^{*}<\mathcal{M}_{\text{PAD}}^{\left\{ 1f\right\} }\simeq0.04$
(see Figs. \ref{fig04-parachor}a and b). In this range close to the
critical point, the Ising-like universal features estimated from the
massive renormalization scheme are then correctly accounted for, in
conformity with a (dimensionless) fluid characterization which only
uses the two scale factors $Y_{c}$ and $Z_{c}$.

\begin{figure}
\includegraphics[width=80mm,keepaspectratio]{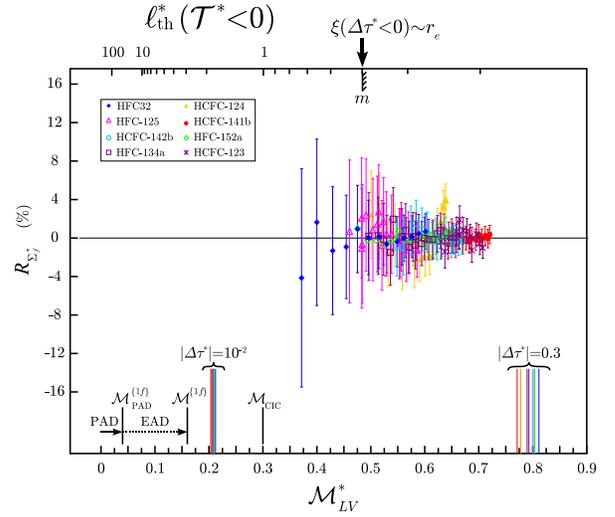}

\caption{(Color online) lin-lin scale: Residuals (expressed in \%) for the
experimental master parachor of eight HFCs and HCFCs (see Refs. \citet{Okada1986,Okada1987,Higashi1992,Okada1995}),
from reference to the master parachor calculated by Eq. (\ref{A11}).\label{fig06}}

\end{figure}

To magnify the relative master behavior at large values of $\mathcal{M}_{LV}^{*}$,
i.e. beyond the preasymptotic range $\mathcal{M}_{LV}^{*}>\mathcal{M}_{\text{PAD}}^{\left\{ 1f\right\} }\simeq0.04$,
we have reported the (\%)-residuals\begin{equation}
R_{\Sigma_{\text{PAD}}^{*}}=100\left[\frac{\Sigma^{*}}{\mathcal{\widetilde{Z}}_{\sigma}\left(\mathcal{M}_{\text{th}}^{*}\right)^{\pi_{a}}}-1\right]\label{A1}\end{equation}
as a function of $\mathcal{M}_{LV}^{*}\equiv\mathcal{M}_{\text{th}}^{*}$
in Fig. \ref{fig05}a, using lin-lin scale. We have also reported
the values $\mathcal{M}_{\text{PAD}}^{\left\{ 1f\right\} }$ {[}see
Eq. (\ref{McalPAD (41)})], $\mathcal{M}^{\left\{ 1f\right\} }$ {[}see
Eq. (\ref{eq:McalEAD (42)})], and $\mathcal{M}_{\text{CIC}}$ {[}see
Eq. (\ref{McalCIC (43)})], which characterize the finite distances
to the critical point where $\ell_{\text{th}}^{*}\simeq40$, $\ell_{\text{th}}^{*}\simeq3$,
and $\ell_{\text{th}}^{*}\simeq1$, respectively (as previously discussed
in § 2.4). This figure \emph{reveals the unambiguous non-universal
nature of each fluid at large distance from the critical point}, i.e.,
when $\ell_{\text{th}}^{*}\lesssim\frac{1}{2}$. Moreover, the observed
master behavior in the extended asymptotic domain $\mathcal{M}_{\text{PAD}}^{\left\{ 1f\right\} }\lesssim\mathcal{M}_{LV}^{*}<\mathcal{M}^{\left\{ 1f\right\} }\simeq0.16$,
i.e., $40\gtrsim\ell_{\text{th}}^{*}\gtrsim3$, needs to introduce
a master correction which can reach $-(10-20)\%$ at the largest extension
$\mathcal{M}^{\left\{ 1f\right\} }\simeq0.16$.

To account for this correction in a quantitative manner, we have used
the following convenient form,\begin{equation}
R_{\Sigma_{\text{mas}}^{*}}=-100A_{\Sigma}\exp\left[-B_{\Sigma}\left(\frac{1}{\mathcal{M}_{\text{th}}^{*}}-\ln\left[\mathcal{M}_{\text{th}}^{*}\right]\right)\right]\label{A2}\end{equation}
which has two adjustable master parameters ($A_{\Sigma}$ and $B_{\Sigma}$)
to control the following main features:

i) within the Ising-like preasymptotic domain, this function takes
zero-value which preserves the Ising-like universal features contained
in $\mathcal{M}_{\text{th}}^{*}\left(\left|\mathcal{T}^{*}\right|\right)$
, only characterized by $Y_{c}$ and $Z_{c}$;

ii) beyond the Ising-like preasymptotic domain, this function must
provide an easy control of supplementary adjustable parameters needed
to recover each real fluid behavior at large distance from the critical
point.

Practically, the master values $A_{\Sigma}=0.575243$ and $B_{\Sigma}=0.302034$
of the adjustable parameters were thus obtained by optimizing the
fit on the experimental \%-residuals of the normal fluids Xe, Kr,
N$_{2}$, and O$_{2}$, as illustrated by the corresponding black
full curve in Fig. \ref{fig05}a. For any other selected fluid, it
was then easy to estimate the related {}``non-universal'' deviation,
expressed in \%, i.e., $\delta R_{\Sigma^{*}}=100\left[\frac{R_{\Sigma_{\text{PAD}}^{*}}}{R_{\Sigma_{\text{mas}}^{*}}}-1\right]$,
as shown in Fig. \ref{fig05}b. We observe that, the larger the value
of the renormalized order parameter density, the more the fluid behaviors
differentiate and the {}``larger'' is the deviation (in amplitude
and shape) from the master crossover behavior.

\begin{table*}
\begin{tabular}{|c|c|c|c|c|c|c|c|c|c|c|}
\hline 
Fluid & $\varphi_{e}$ & $S_{0,e}$ & $\beta_{e}$ & $B_{e}$ & $\phi_{e}$ & $\sigma_{0,e}$ & $\pi_{a,e}$ & $\Sigma_{0,e}$ & $p_{f}$ & $\Sigma_{f}^{*}$\tabularnewline
 &  & ($mm^{2}$) &  &  &  & ($mN\, m^{-1}$) &  & ($mN\, m^{-1}$) &  & \tabularnewline
\hline
\hline 
HFC-32 & $0.935257$ & $8.88119$ & $0.330003$ & $2.00938$ & $1.26526$ & $74.0278$ & $3.833665$ & $5.098332$ & $1,75798$ & $0,145884$\tabularnewline
HFC-125 & $0.9220591$ & $4.985323$ & $0.3182363$ & $1.897228$ & $1.240295$ & $52.68437$ & $3,896303$ & $4.342498$ & $5,014323$ & $0,5406810$\tabularnewline
HFC-134a & $0.8912903$ & $5.830025$ & $0.3115459$ & $1.876448$ & $1.202836$ & $50.00357$ & $3.859691$ & $4.843207$ & $1,087740$ & $0,1497465$\tabularnewline
HFC-152a & $0.904293$ & $8.575218$ & $0.316563$ & $1.903243$ & $1.220856$ & $59.05906$ & $3.856065$ & $4.936506$ & $1,043043$ & $0,1492084$\tabularnewline
HCFC-123 & $0.9785701$ & $6.054669$ & $0.2967141$ & $1.795946$ & $1.275284$ & $59.07641$ & $4.298258$ & $4.769023$ & $2,923514$ & $0,4197302$\tabularnewline
HCFC-124 & $0.9073108$ & $5.270590$ & $0.289472$ & $1.768259$ & $1.196783$ & $51.72997$ & $4,116035$ & $4.912059$ & $2,015201$ & $0,3815439$\tabularnewline
HCFC-141b & $0.9246801$ & $7.477146$ & $0.3048532$ & $1.794835$ & $1.229533$ & $60.5395$ & $4,033156$ & $5.721538$ & $1,678934$ & $0,1952588$\tabularnewline
HCFC-142b & $0.9173884$ & $7.002359$ & $0.3062231$ & $1.903243$ & $1.223612$ & $55.82877$ & $3.996039$ & $5.114294$ & $1,939504$ & $0,2315527$\tabularnewline
 &  &  &  &  &  &  &  &  &  & \tabularnewline
\hline
\end{tabular}

\caption{Fitting values of the effective exponent-amplitude parameters for
power law description by Eqs. (\ref{A3}), (\ref{A5}), (\ref{A6}),
(\ref{A7}), and (\ref{A11}) of experimental interfacial properties
(see Refs. \citet{Okada1986,Okada1987,Higashi1992,Okada1995}) of
eight HFCs and HCFCs. \label{tab3}}

\end{table*}

The theoretical estimation of these fluid differences, for example
in the range of the VLE line which includes the temperature value
$T=0.7\, T_{c}$ where the acentric factor is defined, is out of the
present understanding of any fluid theory. However, we can use a practical
approach detailed in Ref. \citet{LeNeindre2007}, where the main objective
is to recover consistency with the usual description of the surface
tension at large distance from the critical point by the equation\begin{equation}
\sigma=\sigma_{0,e}\left|\Delta\tau^{*}\right|^{\phi_{e}}\label{A3}\end{equation}
In the above effective power law, $\sigma_{0,e}$ and $\phi_{e}$
are the two adjustable parameters which characterize each one-component
fluid. Indeed, the noticeable result is a quasi-constant value of
$\phi_{e}$ which is found in the $1.2-1.3$ range for most of the
fluids, but which significantly differs from the mean-field value
$\phi_{MF}=\frac{3}{2}$, as mentioned in our introduction. In this
Appendix, we limit the analysis to the $S_{g}$ and $\sigma$ data
which are tabulated as a function of $T$ data in Refs. \citet{Higashi1992,Okada1995},
for eight HCFCs and HFCs listed in Table \ref{tab1} (for the complete
study see Ref. \citet{LeNeindre2007}). As in the large number of
experimental works related to the determination of the surface tension,
only the squared capillary length data were effectively measured from
differential capillary rise method (DCRM), as a function of temperature.
Then the surface tension data have been calculated using Eq. (\ref{force balance in g field (6)}),
where the density difference $\rho_{L}-\rho_{V}$ was estimated from
published ancillary equations for the liquid density and vapor density,
along the VLE line. Therefore, introducing the critical temperature
$T_{c}$ and the critical density $\rho_{c}$ of each selected fluid
given in Table, we can re-calculate the $\Delta\rho_{LV}^{*}$ data
at each tabulated value of $T_{c}-T$, using thus the equation\begin{equation}
\Delta\rho_{LV}^{*}=\frac{\sigma}{g\rho_{c}S_{g}}\label{A4}\end{equation}
where $g=9,80665\,\text{m}\text{s}^{-2}$ is the Earth's gravitational
acceleration.

We have fitted the tabulated $\sigma$ data, using Eq. (\ref{A3}).
In addition we have also fitted the tabulated $S_{g}$ and recalculated
$\Delta\rho_{LV}^{*}$ data, using the following equations\begin{equation}
S_{g}=S_{0,e}\left|\Delta\tau^{*}\right|^{\varphi_{e}}\label{A5}\end{equation}
\begin{equation}
\Delta\rho_{LV}^{*}=B_{e}\left|\Delta\tau^{*}\right|^{\beta_{e}}\label{A6}\end{equation}
Thus, we have checked the consistency of the above results, by fitting
the tabulated $\sigma$ data as a function of the recalculated $\Delta\rho_{LV}^{*}$
data, at the same tabulated $T_{c}-T$ data, using the following {}``parachor''
equation\begin{equation}
\sigma=\Sigma_{0,e}\left(\Delta\rho_{LV}^{*}\right)^{\pi_{a,e}}\label{A7}\end{equation}
The adjustable values of $\sigma_{0,e}$, $\phi_{e}$, $S_{0,e}$,
$\varphi_{e}$, $B_{e}$, $\beta_{e}$, $\Sigma_{0,e}$, and $\pi_{a,e}$,
are reported in Table \ref{tab3} and permit to validate the interrelations
$\beta_{e}=\phi_{e}-\varphi_{e}$, $\pi_{a,e}=\left(1-\frac{\varphi_{e}}{\phi_{e}}\right)^{-1}$,
$B_{e}=\frac{\sigma_{0,e}}{g\rho_{c}S_{0,e}}$, and $\Sigma_{0,e}=\frac{g\rho_{c}S_{0,e}}{\left(B_{e}\right)^{\frac{\varphi_{e}}{\beta_{e}}}}$.
As a conclusive remark, it appears that the Eqs. (\ref{A3}) and (\ref{A7})
are explicit results due to the initial use of {}``power laws''
{[}see Eqs. (\ref{A5}) and (\ref{A6})] to fit the {}``measured''
$S_{g}$ data and $\Delta\rho_{LV}^{*}$ data, respectively, in the
same restricted temperature range. The interfacial properties $S_{g}$,
$\rho_{L}-\rho_{V}$, and $\sigma$, of each fluid are then characterized
by two amplitude-exponent pairs, i.e., four adjustable parameters
(in addition to the needed critical parameters which characterize
the liquid-vapor critical point). Moreover, in such a self-consistent
result of the fitting, the relative error-bar on the surface tension
data can then be readily approximated by the sum of the relative error-bars
on the squared capillary length and coexisting relative density measurements
(thus including the relative uncertainty on the $\rho_{c}$ value,
generally of the order of $1\%$). 

In the next step, at each tabulated temperature, we have estimated:

1) the renormalized surface tension\begin{equation}
\Sigma_{\text{tab}}^{*}=\left(\alpha_{c}\right)^{d-1}\beta_{c}\sigma\label{A8}\end{equation}
using the tabulated $\sigma$ data, and

2) the master surface tension\begin{equation}
\begin{array}{cl}
\Sigma_{\text{mas}}^{*} & =\mathcal{\widetilde{Z}}_{\sigma}\left[\left(Z_{c}\right)^{\frac{1}{2}}\Delta\rho_{LV}^{*}\right]^{\pi_{a}}\left\{ 1-\right.\\
 & \left.A_{\Sigma}\exp\left[-B_{\Sigma}\left(\frac{1}{\left(Z_{c}\right)^{\frac{1}{2}}\Delta\rho_{LV}^{*}}-\ln\left[\left(Z_{c}\right)^{\frac{1}{2}}\Delta\rho_{LV}^{*}\right]\right)\right]\right\} \end{array}\label{A9}\end{equation}
using the recalculated $\Delta\rho_{LV}^{*}$ data. The \%-deviations
$\delta R_{\Sigma_{\text{mas}}}=100\left[\frac{\Sigma_{\text{tab}}^{*}}{\Sigma_{\text{mas}}^{*}}-1\right]$
at each value $\mathcal{M}_{LV}^{*}=\left(Z_{c}\right)^{\frac{1}{2}}\Delta\rho_{LV}^{*}$,
have been fitted using the following power law\begin{equation}
\delta R_{\Sigma_{\text{mas}}}=100\Sigma_{f}^{*}\left(\mathcal{M}_{LV}^{*}\right)^{p_{f}}\label{A10}\end{equation}
where $\Sigma_{f}^{*}$ and $p_{f}$ are two adjustable parameters,
given in Table \ref{tab3}. As a final result, the parachor correlation
now reads\begin{equation}
\begin{array}{cl}
\Sigma^{*}= & \mathcal{\widetilde{Z}}_{\sigma}\left(\mathcal{M}_{LV}^{*}\right)^{\frac{\beta}{\phi}}\\
 & \left[1-A_{\Sigma}\exp\left[-B_{\Sigma}\left(\frac{1}{\mathcal{M}_{LV}^{*}}-\ln\left[\mathcal{M}_{LV}^{*}\right]\right)\right]\right.\\
 & \left.+\Sigma_{f}^{*}\left(\mathcal{M}_{LV}^{*}\right)^{p_{f}}\right]\end{array}\label{A11}\end{equation}
where the {}``confluent'' correction to the leading term contains
two parts: i) a master contribution which is the same for all the
pure fluids; ii) a non-universal contribution which is characterized
by the exponent-amplitude pair $p_{f};\Sigma_{f}$ for each pure fluid.
Figure \ref{fig06} gives the corresponding residuals $R_{\Sigma_{f}^{*}}=100\left[\frac{\Sigma_{\text{tab}}^{*}}{\Sigma^{*}}-1\right]$
over the temperature range of the VLE line covered by the capillary
rise measurements. In this figure, we have also reported each error-bar
contribution of the experimental accuracy of $\pm0.2\,\text{mN}\,\text{m}^{-1}$
claimed by the authors (ignoring the corresponding contribution of
the claimed accuracy ($\pm20\,\text{mK}$) on temperature measurements). 

We conclude that the above practical approach based on the correct
master description of the asymptotic Ising-like domain for the one-component
fluid subclass, provides an easy control of the adjustable parameters
needed to recover the experimental behavior at large distance from
the critical temperature.


\begin{thebibliography}{10}
\bibitem{hyperscaling}$\eta$ is the (Fisher) exponent associated
to the universal feature of the correlation function at the exact
critical point ($\Delta\tau^{*}=0$; $\Delta\rho^{*}=0$), while $\delta$
is the exponent associated to the universal feature of the ordering
field along the critical isotherm ($\Delta\tau^{*}=0$; $\Delta\rho^{*}\gtrless0$).
\end{thebibliography}
\end{document}